\documentclass[11pt]{article}

\usepackage{tikz,ifthen,calc}

\usepackage{mathrsfs}           
\usepackage{vmargin,fancyhdr}   
\usepackage{algorithm}
\usepackage{algorithmic}

\usepackage{enumerate}

\usepackage{amsmath,amssymb}    
\usepackage{verbatim}           
\usepackage{xspace}             
\usepackage{graphicx,float}     
\usepackage{ifthen,calc}        
\usepackage{textcomp}           
\usepackage{fancybox}           
\usepackage{hhline}             
\usepackage{float}              

\usepackage[rflt]{floatflt}     
\usepackage[small,compact]{titlesec}
\usepackage{setspace}
\usepackage{subfigure}

\usepackage{color}
\definecolor{ForestGreen}{rgb}{0.1333,0.5451,0.1333}
\usepackage[dvips,
            colorlinks,linkcolor=ForestGreen,citecolor=ForestGreen,
            backref,
            bookmarks,bookmarksopen,bookmarksnumbered]
           {hyperref}
\usepackage[dvips]{thumbpdf}

\setpapersize{USletter}
\setmarginsrb{.75in}{.5in}        
             {0.75in}{.5in}        
             {.25in}{.25in}     
             {.25in}{.5in}      
\newcommand{\showccc}[0]{0}
\newcommand{\ccc}[2][nothing]{
  \ifthenelse{\showccc=0}{}{
    \ensuremath{^{\Lsh\Rsh}}\marginpar{\raggedright\tiny\textsf{%
        \ifthenelse{\equal{#1}{nothing}}{}{\textbf{#1}\\}#2}}}}

\newcounter{hours}\newcounter{minutes}
\newcommand{\hhmm}{%
  \setcounter{hours}{\time/60}%
  \setcounter{minutes}{\time-\value{hours}*60}%
  \ifthenelse{\value{hours}<10}{0}{}\thehours:%
  \ifthenelse{\value{minutes}<10}{0}{}\theminutes}
\ifthenelse{\showccc=0}{\rhead{}}{\rhead{\today \ [\hhmm]}}
\newtheorem{theorem}{Theorem}[section]

\newtheorem{Definition}[theorem]{Definition}

\newtheorem{lemma}[theorem]{Lemma}
\newtheorem{fact}[theorem]{Fact}

\newcommand{\Proof}[0]{\smallskip\noindent\textit{\textbf{Proof}}\quad}
\newcommand{\Proofof}[1]{\smallskip\noindent\textit{\textbf{Proof of #1:}}\quad}

\newcommand{\QED}[0]{\hfill\ensuremath{\blacksquare}\medspace}
\newcommand{\QEDpart}[1]{\hfill\ensuremath{\blacksquare}(#1)\medspace}

\floatstyle{ruled}
\newfloat{algo}{tbp}{lop}
\floatname{algo}{Algorithm}

\usepackage{float}
\floatstyle{plain}\newfloat{myfig}{t}{figs}[section]
\floatname{myfig}{\textsc{Figure}}
\floatstyle{plain}\newfloat{myalg}{H}{algs}[section]
\floatname{myalg}{}
\newcommand{\component}[1]{G^{(#1)}}
\newcommand{\componenth}[1]{H^{(#1)}}
\newcommand{\Gschurapprox}{\tilde{G}_{schur}}

\newcommand{\eflow}{\mathcal{E}}
\newcommand{\epotential}{\mathcal{E}}

\newcommand{\fail}{\textbf{``fail"}}

\newcommand{\approxflow}{\tilde{\textbf{f}}}
\newcommand{\approxflowv}{\tilde{\textbf{f}}}
\newcommand{\optflow}{\bar{\textbf{f}}}
\newcommand{\optflowv}{\bar{\textbf{f}}}
\newcommand{\flow}{\textbf{f}}
\newcommand{\flowv}{\textbf{f}}

\newcommand{\potentialvert}{\phi}
\newcommand{\potentialvertv}{\phi}

\newcommand{\approxpotentialvertv}{\tilde{\phi}}

\newcommand{\approxpotentialschurv}{\tilde{\phi}_{schur}}

\newcommand{\mata}{\textbf{A}}
\newcommand{\matb}{\textbf{B}}
\newcommand{\mati}{\textbf{I}}
\newcommand{\maty}{\textbf{Y}}

\newcommand{\vecx}{\textbf{x}}
\newcommand{\vecxboundary}{\textbf{x}_{bdry}}
\newcommand{\vecxinterior}{\textbf{x}_{intr}}
\newcommand{\vecy}{\textbf{y}}

\newcommand{\vecyx}{
\left(
\begin{array}{c}
\vecy\\
\vecx
\end{array}
\right)
}

\newcommand{\poly}[1]{\textbf{poly}(#1)}

\newcommand{\weightcombined}{\textbf{w}_{combined}}

\newcommand{\weightoracle}{\textbf{w}_{oracle}}
\newcommand{\weightoraclev}{\textbf{w}_{oracle}}
\newcommand{\weight}{\textbf{w}}
\newcommand{\weightv}{\textbf{w}}
\newcommand{\weightd}{\textbf{W}}

\newcommand{\weightgroup}{\textbf{w}_{grp}}
\newcommand{\weightgroupv}{\textbf{w}_{grp}}
\newcommand{\resistance}{\textbf{r}}
\newcommand{\resistancev}{\textbf{r}}

\newcommand{\resistanced}{\textbf{R}}

\newcommand{\edgevertex}{\textbf{B}}

\newcommand{\laplacian}{\textbf{L}}
\newcommand{\laplacianapprox}{\tilde{\textbf{L}}}
\newcommand{\laplacianinner}{\textbf{L}_{intr}}
\newcommand{\laplacianmiddle}{\textbf{L}_{mid}}
\newcommand{\laplacianboundary}{\textbf{L}_{bdry}}
\newcommand{\laplacianpivoted}{\textbf{L}_{schur}}
\newcommand{\laplacianpivotedapprox}{\tilde{\textbf{L}}_{schur}}

\newcommand{\maxcond}{\kappa}
\newcommand{\maxratio}{U}

\newcommand{\congestion}[2]{\textbf{cong}(#1, #2)}

\newcommand{\zerosv}{\textbf{0}}

\newcommand{\capacity}{\textbf{u}}
\newcommand{\capacityv}{\textbf{u}}
\newcommand{\Vboundary}{V_{bdry}}
\newcommand{\Vinterior}{V_{intr})}
\newcommand{\VboundaryG}[1]{V_{bdry}(#1)}
\newcommand{\VinteriorG}[1]{V_{intr}(#1)}

\newcommand{\demand}{\textbf{d}}
\newcommand{\demandv}{\textbf{d}}

\newcommand{\demandschurv}{\textbf{d}_{schur}}

\newcommand{\demandboundary}{\textbf{d}_{bdry}}
\newcommand{\demandboundaryv}{\textbf{d}_{bdry}}

\newcommand{\numiter}{N}

\begin{document}

\title{Approximate Maximum Flow on Separable Undirected Graphs
\thanks{Partially supported by the National Science Foundation under grant number CCF-1018463.}}

\author{
  Gary L.\ Miller\\
  CMU\\
  \texttt{glmiller@cs.cmu.edu}\\
  \and
  Richard Peng \thanks{Supported by a Microsoft Fellowship}\\
  CMU\\
  \texttt{yangp@cs.cmu.edu}\\
}



\maketitle

\begin{abstract}
We present faster algorithms for approximate maximum flow in
undirected graphs with good separator structures, such as bounded
genus, minor free, and geometric graphs.
Given such a graph with $n$ vertices, $m$ edges along with
a recursive $\sqrt{n}$-vertex separator structure,
our algorithm finds an $1-\epsilon$ approximate maximum flow in time
$\tilde{O}(m^{6/5} \poly{\epsilon^{-1}})$,
ignoring poly-logarithmic  terms.
Similar speedups are also achieved for separable graphs with larger size
separators albeit  with larger run times.  These bounds also apply
to image problems in two and three dimensions.

Key to our algorithm is an intermediate problem that we term grouped $L_2$
flow,  which exists between maximum flows and electrical flows.
Our algorithm also makes use of spectral vertex sparsifiers in order to
remove vertices while preserving the energy dissipation of electrical flows.
We also give faster spectral vertex sparsification algorithms on well separated
graphs, which may be of independent interest.
\end{abstract}

\section{Introduction}
\label{sec:intro}

The maximum flow problem is a fundamental algorithmic question introduced
by Ford and Fulkerson \cite{FordF56} and studied extensively since.
In its simplest form, the problem asks for a flow $\flowv$ in
a graph $G$ that routes the maximum amount from a source
$s$ to a sink $t$,
subject to the constraint that the amount of flow on each edge
does not exceed  its capacity.

Due to its origins in studying railroad network problems
\cite{HarrisR56}, the maximum flow problem on planar graphs
has also received considerable attention.
The Ford-Fulkerson paper gives an algorithm for
the case where $s$ and $t$ lie on the same face \cite{FordF56}.
In this setting a long line of work \cite{Hassin81, JohnsonV82, MillerNaor95}
led to a nearly-linear time algorithm on directed planar graphs\cite{BorradaileK09}.
In the case of undirected graphs, which our work addresses,
the running time has been brought to as low as
$O(n\log\log{n})$ \cite{ItalianoGNSW11}.
While originally intended for transportation problems, these algorithms also
have a wide range of applications in computer vision \cite{BorradaileThesis}.

Many applications contain a number of
irregularities that are either intrinsic or added by the user.
These irregularities are sufficiently common to be studied in
early works on planar separators by Lipton and Tarjan \cite{LiptonTarjan79}
(e.g. a 2D finite element graphs where one connects any two vertices that share
a face).
They can take the form of bridges in transportation problems,
or extra correlations in non-local methods for images \cite{buades08}.
As a result, algorithms resilient to a small number of edge
changes are needed to handle many of these instances.
Progress in this direction was made by Chambers et al. \cite{ChambersEN09a,ChambersEN09b}, who gave
an algorithm for $s$-$t$ maximum flow on a genus $g$ graph
with integer capacities summing up to $C$ that
runs in ${O}(\poly{g}n\log^2n\log^2C)$ time.

It is also possible for small modifications to
introduce highly connected minors while preserving
the overall structure of the graph.
For example, two or more layers of an $O(\sqrt{n}) \times O(\sqrt{n})$
grid with corresponding vertices connected on the layers is similar to
the (planar) square grid, yet has a large $K_{\sqrt{n}}$ minor.
Graphs similar to this often occur in computer vision problems
involving 3-D images, or 2-D videos \cite{VastaSN09, LiuFreeman10}.
As a result, we believe that a more robust algorithm can
be directly used for a considerably wider range of applications.

In this work we introduce a novel approach for computing approximate
maximum flow on these graph families.
We show that good separator structures alone are sufficient for faster
algorithms for computing a flow within $1 - \epsilon$ of the maximum flow.
Our work is  based on the multiplicative weights method recently 
used by Christiano et al. \cite{ChristianoKMST10} in their faster
approximate maximum flow algorithm.
We develop a two-level, recursive approach that gives
even faster algorithms on separable undirected graphs.
To our knowledge, the only previous work using just
separator structures was by Imai and Iwano \cite{ImaiI90}.
Their algorithm takes advantage of the separators by speeding up
the $L_2$ minimizations, giving a running time of about ${O}(m^{1.59})$.
As a maximum flow algorithm, however, their result has been superceded by
the result on general graphs by Goldberg and Rao \cite{GoldbergR98}.

The Christiano et al. algorithm uses very different techniques
than previous work.
Their algorithm constructs an approximate maximum flow
by taking the average of a sequence of electrical flows,
each constructed based on previous ones.
These electrical flows are in turn computed using the nearly-linear
time solvers for SDD linear systems introduced by Spielman
and Teng \cite{SpielmanTeng04, SpielmanTengSolver, KoutisMP10, KoutisMP11}.
The solver algorithm can be viewed as recursion on a
hierarchy of graphs, each increasingly sparser than the previous one.
Direct adaptions of this type of recursion on gradually shrinking
graphs have yielded fast approximate cut algorithms \cite{Madry10b},
albeit at the cost of some loss in the quality of the solution.
This work naturally leads to the question of whether the Christiano et al.
algorithm can be combined with some type of recursive routine
to get faster algorithms with smaller errors.

In a work joint with Hui Han Chin and Aleksander M\c{a}dry
\cite{ChinMMP12} that addressed problems from image processing,
we observed that a coarser grouping of edges leads to an intermediate
problem between maximum flow and electrical flow.
This problem, which we term the \textbf{grouped $L_2$ flow} problem,
has two key properties that we'll show.
It can be used in place of electrical flows in the Christiano et al.
algorithm and lead to fewer iterations when the groups are small.
Furthermore, on well separated graphs, these problems
can be approximated in nearly-linear time for intermediate values of $k$.
Combining the decreased iteration count with the nearly-linear running
time of each iteration gives a faster algorithm.
Our algorithm can also be readily extended to many of the
image processing related tasks, such as isotropic total
variation minimization \cite{GoldfarbY04}.
Due to the additional components of these objectives and
the readiness of the extension, we restrict our presentation
to only approximating the maximum flow.

Our algorithm constructs the intermediate grouped flow problem based
on a $r$-division, which are partitions of the graph into $n/r$ groups
each with small boundary of size $O(\sqrt{r})$.
Then we show that vertices incident to edges from only one
group can be removed using \textbf{spectral vertex sparsifiers}.
This removal obtained by modifying the generalized nested
dissection algorithm by Lipton, Rose and Tarjan \cite{LiptonRT79}.
This result allows us to reduce the problem size by a factor of $\Omega(\sqrt{r})$,
leading to savings in the cost of computing these flows.
When given the separator structures generated by algorithms
such as \cite{LiptonTarjan79,KawarabayashiR10,Wulffnilsen11},
our main result gives:

\begin{theorem}
\label{thm:main}
Given a graph $G = (V, E)$ with $n$ vertices and $m$ edges
with vertices partitioned into a $r$-division along with recursive
separator trees for each group,
capacities $\capacity \in \Re^{E}$ and source/sink vertices $s$ and $t$,
we can compute a $1 - \epsilon$ approximate maximum $s-t$ flow in
$\tilde{O}(m^{6/5}\epsilon^{-6.1})$ \footnote{We use $\tilde{O}(f(m))$ to denote
$\tilde{O}(f(m) \log^c f(m))$ for some constant $c$.}  time.
\end{theorem}

Our results are applicable to a variety of graph families where
separators can be found efficiently.
Families  where nearly-linear time separator algorithms are known
include planar \cite{LiptonTarjan79}, bounded genus, $h$-minor
free for fixed $h$ \cite{KawarabayashiR10} and nearest neighbor
geometric graphs \cite{MiTeThVa97a}.
For $h$-minor free graphs with larger values of $h$ (e.g. $n^{\epsilon}$),
speedups can still be obtained using the algorithm by Wulff-Nilsen
\cite{Wulffnilsen11}, which runs in $\tilde{O}(n^{5/4+O(\epsilon)})$ time.

Aside from giving the first speedup over approximate general flow
algorithms for the class of minor-free and shallow minor-free
graphs, our weaker requirement of separator structure have
several other advantages.
The requirement of a top-level partition into $n/r$ pieces 
naturally incorporate $O(n/r)$ extra edges.
When an even larger number of such extra edges are present,
the performance of our algorithm can degrade smoothly to that of 
Christiano et al. algorithm by a suitable change of parameters.
Similar modifications can also be done if the separators
given are of size $n^{\beta}$ instead of $\sqrt{n}$.
If $\beta < 2/3$, a modified version of our algorithm still
gives running times faster than $\tilde{O}(m^{4/3}\poly{\epsilon^{-1}})$.

This robustness of our algorithm means it extends
naturally to multiple sources and sinks,
which occur in a variety of applications.
On general graphs, this can be handled by introducing a super
source and super sink, then connecting all vertices to them
with appropriate capacities.
However, this modification can significantly change
the connectivity structure.
As a result this extension on planar graphs has been studied as
an important problem on its own \cite{MillerNaor95},
with a nearly-linear time algorithm being a recent result \cite{BorradaileKMNW11}.
The milder requirement of partitions and separators on the other
hand can naturally incorporate a super source and sink.
These two vertices can be incorporated as part of every
partition with only additive constant overhead.
This means the guarantees in Theorem \ref{thm:main} includes
the multiple sources/sinks case, and may have other advantages
in terms of robustness.

Our results can also be combined with graph smoothing and
sampling techniques \cite{Karger98} to obtain speedups by a
factor of about $(m/n)^{1/5}$.
As planar, bounded genus, and minor-free graphs
have $m = O(n)$,  this improvement is only applicable 
for dense graphs that have good underlying
vertex separator structures, and we omit it from our presentation.

The paper is organized as follows: Section \ref{sec:background} gives
formal definitions of the tools  we use.
The main components of our algorithm and their
interactions are described in Section \ref{sec:overview}.
The analysis of these components are direct adaptations of
known results, with error analyses incorporated appropriately.
They are given in Sections \ref{sec:approxgrouped},
\ref{sec:conversion} as well as Appendices
\ref{sec:groupedflow}, \ref{sec:vertexsparsify}.
Possible extensions of our results are discussed in
Section \ref{sec:comments}.

\section{Background and Notations}
\label{sec:background}

In this section we review some of concepts and notation used in the
paper.
We use subscripts to distinguish between variants of variables
and $(\cdot)$ to denote entries in vectors/matrices.

\subsection{Graphs, Graph Laplacians and Spectral Ordering of Matrices}

We use $G = (V, E)$ to denote an undirected graph.
If we orient each edge $e=uv$ arbitrarily,
the edge-vertex incidence matrix of $G$ is defined as:
\begin{align*}
\edgevertex(e, u) =
\left\{
\begin{array}{lr}
-1 & \text{if u is the head of e} \\
1 & \text{if u is the tail of e} \\
0 & \text{otherwise}
\end{array}
\right.
\end{align*}

Given edge weights of $G$, $\weightv$, we can define
the Laplacian of $G$ as $\laplacian = \edgevertex^T \weightd
\edgevertex$, where $\weightv$ is the diagonal matrix of edge weights.
Entry-wise we have that $\laplacian(u, v)$ is the degree of vertex $u$
if $v = u$; $-\weight_{uv}$ if $uv$ is an edge; and $0$ otherwise.
Many of our routines have logarithmic dependency on the
ratio between the maximum and minimum entry of $\weightv$.
In all of their invocations needed for our main result,
this parameter is bounded by $\poly{n}$.
As a result, this term can be viewed as an additional factor of
$\log{n}$ and is not crucial to understanding our approach.
In the formal presentation of our results, we will use
$\maxratio(\weightv)$ to denote $\max(\weightv) / \min(\weightv)$,
and will also apply this notation to other vectors.

We will use $\laplacian$ interchangeably with $G$.
When the underlying graph is not clear from the context,
we will also denote it using $\laplacian(G)$.
It can be shown that graph Laplacians are positive semi-definite.
A standard notation that we'll use repeatedly throughout our presentation
for comparing matrices is $\preceq$.
Specifically,  if $\mata$ and $\matb$ are positive semi-definite matrices, we
use $\mata \preceq \matb$ to denote that $\matb - \mata$ is positive
semi-definite.
Or in other words, $\vecx^T \mata \vecx \leq \vecx^T \matb \vecx$ for
all vector $\vecx$.
We will also use $0 = \lambda_{1}(\laplacian) \leq \lambda_{2}(\laplacian) 
\leq \ldots \leq \lambda_{n}(\laplacian)$ to denote the $n$ eigenvalues
of $\laplacian$.

\subsection{Approximate Maximum Flow}

If we're also given capacities on the edges of $G$,
$\capacityv: E \rightarrow \Re^+$.
Then for a flow on $G$, $\flowv: E \rightarrow \Re^+$,
we can define the (undirected) \textbf{congestion} of this flow
w.r.t. edge $e$ as:
\begin{align*}
\congestion{\flowv}{e} = \frac{|\flowv(e)|}{\capacity(e)}
\end{align*}

In an undirected graph, the requirement that a flow obeys capacities
can be expressed as $\congestion{\flowv}{e} \leq 1$.
We would also like the flow to meet certain demands at vertices.
Specifically given a demand vector $\demandv$ that sums to $0$,
this constraint can be written as $\edgevertex^T \flowv = \demandv$.
Since we're allowed to add a super source/sink, one demand vector of
particular interest is $\demandv$ being only non-zero at two vertices, $s$ and $t$.
This corresponds to the $s$-$t$ maximum flow problem,
which seeks to maximize the amount of flow entering $t$
(or equivalently, exiting $s$) while obeying capacities and
routing the same amount in and out of all other vertices.
The approximate maximum flow problem in turn becomes finding a flow
that routes at least $1 - \epsilon$ of the maximum amount.
As flows on connected components behave independently of each other,
we may also assume that $G$ is connected.

\subsection{Electrical Flows}

Given a set of resistive values on the edges,
$\resistancev : E \rightarrow \Re^+$, the electrical energy dissipation
of a flow $\flowv$, $\eflow_\flowv(\resistancev)$, can be defined as:
\begin{align*}
\eflow_\flowv(\resistancev)= \sum_e \resistance(e) \flow(e)^2
\end{align*}

The set of resistances also defines a natural graph Laplacian, namely
$\laplacian = \edgevertex^T \resistanced^{-1} \edgevertex$.
For a demand $\demandv$, let $\optflowv$ denote the flow that minimizes
the electrical energy and let $\eflow(\resistancev) = \eflow_{\optflowv}(\resistancev)$.
We use the following facts,  shown in Section 2.3.1 of \cite{ChristianoKMST10},
about the energy dissipation  of the optimum electrical flow.

\begin{lemma}
\label{lem:optelectrical}
\begin{align*}
\eflow(\resistancev) = \demandv^T \laplacian^{+} \demandv
\end{align*}
Where $\demandv$ is the column demand vector,
and $\laplacian^{+}$ is the pseudo-inverse of $\laplacian$.
\end{lemma}

A more thorough treatment of various properties of electrical flows and
associated voltage values can be found in \cite{doylesnell84}.
In order to compute electrical flows, we invoke the following result about computing
almost-optimal electrical flows using SDD linear system solvers 

\begin{theorem}
\label{thm:approxelectrical}
There is an algorithm \textsc{ElectricalFlow} such that when given a weighted
graph $G = (V, E, \resistancev)$,
along with any demand vector $\demandv$ such that the corresponding minimum energy
flow is $\optflowv$.
\textsc{ElectricalFlow} computes in $\tilde{O}(m
\log(\maxratio(\resistancev)  \delta^{-1}))$
time a flow $\approxflowv$ such that

\begin{enumerate}
\item
\label{part:conservationofflow}
$\approxflowv$ meets the demands
$\edgevertex^T \approxflowv = \demandv$
\item
\label{part:flowenergy}
$
\eflow(\approxflowv) \leq (1+\delta) \eflow(\resistancev)
$
\item
\label{part:flowdifference}
For every edge $e$,
$
\sum_e |\resistance(e) \optflow(e)^2  - \resistance(e) \approxflow(e)^2| \leq \delta \eflow_{\optflow}(\resistancev)
$
\end{enumerate}
\end{theorem}

A more detailed discussion of electrical flows, and its
natural dual, effective resistances,
can be found in Section 2 of \cite{ChristianoKMST10}.

\subsection{Graph Separators}

Our algorithm makes extensive use of separator structures, both as
$r$-divisions and as recursive separator trees.
A single vertex separator can be formalized as:

\begin{Definition}
\label{dfn:separator}

On a graph $G$ with vertex set $V$, a subgraph $S$ is an
$\alpha$-separator if $G \setminus S$ can be partitioned into two
graphs $\component{1}$ and $\component{2}$ such that there are no
edges between them and $|V(\component{1})|, |V(\component{2})| \leq
\alpha|V(G)|$.
\end{Definition}

There is a rich literature on efficiently finding such separators for planar,
bounded genus, minor free, and geometric graphs \cite{LiptonTarjan79,
MiTeThVa97a, KawarabayashiR10, Wulffnilsen11}.
Our algorithms do not change the connectivity of the graph,
and use the separators repeatedly.
As a result, we treat structures related to separators as
separate entities computed before running our algorithms.

Given a subgraph $G'$ of $G$, we let its \textbf{vertex boundary}
 $\Vboundary(G')$ be vertices that have edges to $V \setminus V(G')$ and denote
$V(G') \setminus \VboundaryG{G'}$ as $\VinteriorG{G'}$.
We can now define $r$-divisions, which are multi-way partitions
of $V(G)$.

\begin{Definition}
\label{dfn:rdivision}
A $r$-division of a graph $G = (V, E)$ is a partition of it into groups
$\component{1} \ldots \component{k}$ such that each edge belongs to
exactly one group and:
\begin{itemize}
    \item The number of groups, $k$ is at most $O(\frac{n}{r})$
    \item $|E(\component{i})| \leq r$
    \item The number of boundary vertices of each group,
        $\VboundaryG{\component{i}}$ has size at most $O(\sqrt{r})$.
\end{itemize}
\end{Definition}

Fredrickson \cite{Fre87} showed that with an extra $\log{n}$ factor in
running time, one could compute separators repeatedly to obtain
a $r$-division of the graph into pieces of size $O(r)$.
In the planar case this was further improved to linear time by
Goodrich \cite{Goodrich95}.

One of our algorithms for finding sparser equivalents of each
group makes use of a recursive separator trees.
These are obtained by recursively computing separators on
the components, leading to a hierarchical partitioning of the graph.
These structures were first used in nested-dissection algorithms
 \cite{GilbertT87, LiptonRT79}.
A subtle, but significant distinction exists in their construction.
For planar, or more generally sparse-contractible graphs, it was shown
that building separators on each component suffices
for a fast algorithm \cite{GilbertT87}.
However, if we are to only assume separator based properties, we need
also keep the separator itself in one of the components.
This ensures that the separator itself is partitioned in future
layers, and won't eventually become a bottleneck in the algorithm.
Lipton, Rose and Tarjan \cite{LiptonRT79} showed in their
generalized nested dissection algorithm that these
more carefully constructed separator trees alone are
sufficient for good guarantees.
The separators trees that we rely on are based on
their construction and is described below.

\begin{Definition}
A recursive $\alpha$-separator structure $\mathcal{S}(\bar{G})$
for a graph $\bar{G} = (V, E)$ is
a tree where each node corresponds to a subgraph and
satisfies the following properties:

\begin{itemize}
    \item The root node equals to $\bar{G}$.
    \item Each leaf node $G$ has $|V(G)| \leq C$,
         where $C$ is an absolute constant.
    \item Each non-leaf node corresponding to graph $G$ has
         a $\alpha$-separator $S$.
         Furthermore its two children corresponds to
         graphs induced on $\component{1} \cup S$ and
         $\component{2} \cup S$ with edges in $S$ belonging to both.
\end{itemize}
\end{Definition}

Theorem 1 of \cite{LiptonRT79} showed that finding the full
separator tree can also be done with an extra $\log{n}$ factor
overhead in the running times of computing separators.

\subsection{Schur Complement}
\label{subsec:schurcomplement}

The partition of $V$ into $\Vinterior$ and $\Vboundary$ naturally allows us
to partition $\laplacian$ into 4 blocks:
\begin{align*}
\laplacian = \left [
\begin{array}{cc}
\laplacianinner & \laplacianmiddle \\
\laplacianmiddle^T & \laplacianboundary\\
\end{array}
\right]
\end{align*}
Note that by our assumption of $G$ being connected,
$\laplacianinner$ is invertible when both $\Vboundary$ and $\Vinterior$
are non-empty.
The Schur complement acts as a smaller-sized equivalent of $\laplacian$
on only the boundary vertices.

\begin{Definition}
\label{dfn:schurcomplement}
The Schur complement of a graph Laplacian $\laplacian$ with respect to
a set of boundary vertices $\Vboundary$,
$\textsc{SchurComplement}(\laplacian, \Vboundary)$ is:
\begin{align*}
\laplacianboundary - \laplacianmiddle^T \laplacianinner^{-1} \laplacianmiddle
\end{align*}
\end{Definition}

We will also label objects associated with the Schur complement
with the subscript $_{schur}$.
Specifically, the Schur complement itself will also be denoted
using $\laplacianpivoted$.

\section{Overview of Our Algorithms}
\label{sec:overview}

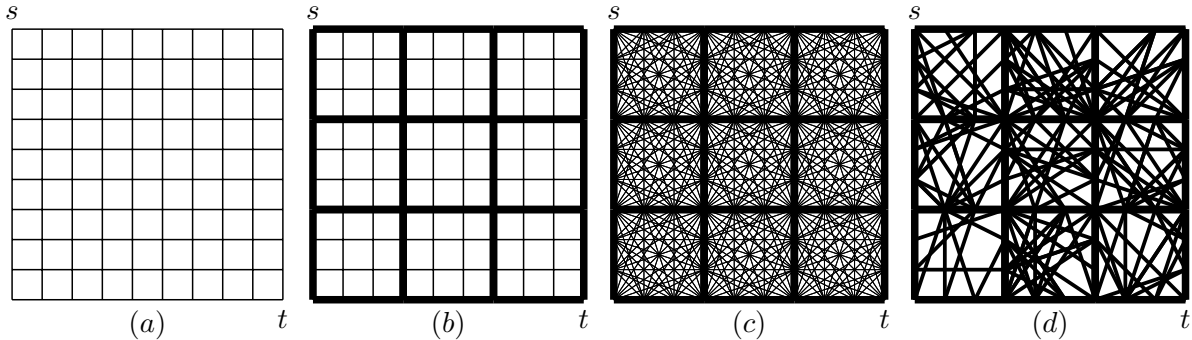
\begin{figure}[ht]
\begin{center}
\begin{tikzpicture}[x=0.4cm, y=0.4cm]

	\begin{scope}[xshift = 0cm]
	 	\draw[line width = 0.2mm] \foreach \x in {0,1,...,9}
		{
			(\x, 0)  -- (\x, 9)
			(0, \x, 0)  -- (9, \x)
		};
		\draw(0,9)node[above]{$s$}; 
		\draw(9,0)node[below]{$t$}; 
		\draw(4.5,0)node[below]{$(a)$}; 
	\end{scope}

	\begin{scope}[xshift = 4cm]
	 	\draw[line width = 0.2mm] \foreach \x in {0,1,...,9}
		{
		        (\x, 0)  -- (\x, 9)
			(0, \x, 0)  -- (9, \x)
		};

		\foreach \xpiece in {0,3, 6} {
		\foreach \ypiece in {0,3, 6} {
			\foreach \small in {0, 3} {
				\draw[line width = 1mm] ({\xpiece + \small}, \ypiece) -- ({\xpiece + \small}, {\ypiece + 3});
				\draw[line width = 1mm] (\xpiece, {\ypiece + \small}) -- ({\xpiece + 3}, {\ypiece + \small});
			}
		}
		}

		\draw(0,9)node[above]{$s$}; 
		\draw(9,0)node[below]{$t$};
		\draw(4.5,0)node[below]{$(b)$}; 
	\end{scope}

	\begin{scope}[xshift = 8cm]
		\foreach \xpiece in {0,3,6} {
		\foreach \ypiece in {0,3,6} {
			\foreach \small in {0, 3} {
				\draw[line width = 1mm] ({\xpiece + \small}, \ypiece) -- ({\xpiece + \small}, {\ypiece + 3});
				\draw[line width = 1mm] (\xpiece, {\ypiece + \small}) -- ({\xpiece + 3}, {\ypiece + \small});
			}
		}
		}

		\foreach \xpiece in {0,3,6} {
		\foreach \ypiece in {0,3,6} {
			\foreach \smallax/\smallay in {0/0,0/1,0/2,0/3,3/0,3/1,3/2,3/3,1/0,2/0,1/3,2/3} {
			\foreach \smallbx/\smallby in {0/0,0/1,0/2,0/3,3/0,3/1,3/2,3/3,1/0,2/0,1/3,2/3}  {
				\draw[line width = 0.1mm] ({\xpiece + \smallax}, {\ypiece + \smallay}) -- ({\xpiece + \smallbx}, {\ypiece + \smallby});
			}
			}
		}
		}

		\draw(0,9)node[above]{$s$}; 
		\draw(9,0)node[below]{$t$};
		\draw(4.5,0)node[below]{$(c)$}; 
	\end{scope}

	\begin{scope}[xshift = 12cm]
		\foreach \xpiece in {0,3,6} {
		\foreach \ypiece in {0,3,6} {
			\foreach \small in {0, 3} {
				\draw[line width = 1mm] ({\xpiece + \small}, \ypiece) -- ({\xpiece + \small}, {\ypiece + 3});
				\draw[line width = 1mm] (\xpiece, {\ypiece + \small}) -- ({\xpiece + 3}, {\ypiece + \small});
			}
		}
		}

		\pgfmathsetseed{123}
		\foreach \xpiece in {0,3,6} {
		\foreach \ypiece in {0,3,6} {
			\foreach \smallax/\smallay in {0/0,0/1,0/2,0/3,3/0,3/1,3/2,3/3,1/0,2/0,1/3,2/3} {
			\foreach \smallbx/\smallby in {0/0,0/1,0/2,0/3,3/0,3/1,3/2,3/3,1/0,2/0,1/3,2/3}  {
				\pgfmathparse{int(mod(int(rand*1000), 5))} \let\res\pgfmathresult
				\ifthenelse{\res = 0}{\draw[line width = 0.5mm] ({\xpiece + \smallax}, {\ypiece + \smallay}) -- ({\xpiece + \smallbx}, {\ypiece + \smallby});}{}
			}
			}
		}
		}

		\draw(0,9)node[above]{$s$}; 
		\draw(9,0)node[below]{$t$};
		\draw(4.5,0)node[below]{$(d)$}; 
	\end{scope}
\end{tikzpicture}
\end{center}
\caption{Illustrations of the main structures that occur
in our algorithm from left to right
(a) original graph;
(b) grid partitioned into $r$-divisions;
(c) exact dense spectral vertex sparsifier;
(d) sparse spectral vertex sparsifier.}
\label{fig:mesh}
\end{figure}

We present a summary of our results in this section.
The bounds for our algorithms are probabilistic due to calls to
spectral sparsifiers \cite{SpielmanS08} that randomly sample graphs.
It was shown in \cite{KL11} that with constant factor slowdown in running time,
these routines can succeed with high probability.
Therefore we omit the failure probability in the guarantees of our
algorithms to simplify our presentation.
An example of applying our algorithm to an instance on a
square mesh is given at the end of this section, with
illustrations given in Figure \ref{fig:mesh}.
The running time of various components are also given
in Figure \ref{fig:summary} in Section \ref{sec:comments}.
It might be helpful to refer to this example, for intuition on
the purposes of each component of our algorithm.

The Christiano et al. algorithm \cite{ChristianoKMST10} showed that
electrical flows can be combined to form a flow that approximately
meets all the edge capacities.
This connection between electrical and maximum flows can be illustrated
by a closer look at the objective functions of these two flows.
Maximum flow imposes capacity constraints of $\flow(e) \leq \capacity(e)$
on each edge, which are equivalent to $(\flow(e) / \capacity(e)) ^2 \leq 1$.
The weights in the $(\epsilon, \rho)$ oracle, and in turn electrical flows
 can be viewed as a way to
create a single constraint from all $m$ edges, in the form of
$\sum_{e} \weight(e) (\flow(e) / \capacity(e)) ^2 \leq \sum_e \weight(e)$.

Instead of aggregating these into a single constraint, we can aggregate
over $k$ groups of edges, leading to an intermediate problem.
The congestion of an edge can then be generalized to that of a group of
edges, and we define $\congestion{\flowv}{i}$ as the square
root of the total energy among edges in the $i$-th group, $\component{i}$:
\begin{align*}
\congestion{\flowv}{i} = \sqrt{\sum_{e \in E(\component{i})} \weight(e) \flow(e)^2}
\end{align*}

Then given a demand vector $\demandv$, an instance of \textbf{grouped $L_2$ flow}
can be formalized as minimizing the maximum congestion over the $k$
groups.
\begin{Definition}
\label{def:groupedl2flow}
A {\bf grouped $L_2$ flow} problem is, given a graph $G=(V,E)$ and edge
weights $\weightv$, source/sink pair $s$ and $t$,
along with a partition of the edges into $S_1,\ldots, S_k$,
find a flow $\flowv$ that meets the demands,
$\edgevertex^T \flowv = \demandv$.
The congestion of this flow is the maximum congestion over
all groups, $\max_i \congestion{\flowv}{i}$.
\end{Definition}

This problem is also the natural dual of the grouped least squares
problem introduced in \cite{ChinMMP12}, which generalizes
a variety of objective functions from image processing.
We say that an algorithm approximates grouped $L_2$ flows
if given a graph where there exists a flow $\flowv$
with $\congestion{\flowv}{i} \leq 1 - \epsilon$, it returns
a grouped $L_2$ flow with $\congestion{\flowv}{i} \leq 1 + \epsilon$,
and possibly \fail \quad otherwise.
We show in Section \ref{sec:approxgrouped} that if each group
has at most $r$ edges, the approximate maximum flow problem
can be solved using a sequence of about $r^{1/2}$
approximate grouped $L_2$ flow problems.

\begin{lemma}
\label{lem:groupedfloworacle}
Suppose we have access to a routine for approximating grouped $L_2$
flows, $\textsc{ApproxGroupedFlow}$.
If $\component{1}, \component{2} \ldots \component{k}$
satisfy $|\component{i}| \leq r$ and $\epsilon < 1/2$, an $1-\epsilon$
approximate maximum flow can be computed by
$\tilde{O}(r^{1/2} \epsilon^{-1/2})$ calls to
\textsc{ApproxGroupedFlow}.
Furthermore, in each of these calls, the edge weights $\weightv$
satisfy $\maxratio(\weightv) \leq O(m^3 \epsilon^{-3})$.
\end{lemma}

The rest of our algorithm can then be viewed as giving a series
of faster algorithms for approximating grouped $L_2$ flows.
We first show that the group $L_2$ flows
can be approximated in $O(k^{1/3})$ iterations,
each involving a SDD linear system solve.
This is done by modifying the algorithms from
\cite{ChristianoKMST10, ChinMMP12}.
This algorithm and its analysis are described in Appendix \ref{sec:groupedflow}.
Note that the number of iterations only depends on $k$, the number of groups.
Also, this algorithm returns \fail when it can certify
that a grouped flow with small congestion does not exist.
As a result, our approximate max-flow algorithm, as
stated, only produces a flow.
We show in Appendix \ref{sec:cut} that it can be modified
to return a cut of small value instead.

\begin{theorem}
\label{thm:groupedflow}
Given a graph $G = (V, E, \weightv)$ with edges partitioned
into $k$ groups $S_1 \ldots S_k$,
along with an edge weight vector $\weightv : E \rightarrow \Re^+$
and a demand vector $\demandv: V \rightarrow \Re$.
If there exists a flow $\optflowv$ that meets the demands and satisfies
$\congestion{\optflowv}{i} \leq 1 - \epsilon$, then
$\textsc{GroupedFlow}(G, \demandv, \weightv, \epsilon/10)$
returns in $\tilde{O}(mk^{1/3}\log(\maxratio(\weightv)) \epsilon^{-8/3})$ time
 a flow $\approxflowv$ where for all groups we have:
$\congestion{\approxflowv}{i} \leq (1+\epsilon)$.
\end{theorem}

We can combine this algorithm directly with
Lemma \ref{lem:groupedfloworacle} by grouping
edges into $k = O(m/r)$ groups of size $r$.
This leads to a total running time of:
\begin{align*}
& \tilde{O}(mk^{1/3}\epsilon^{-8/3}) \cdot \tilde{O}(r^{1/2}\epsilon^{-5/2}) \nonumber \\
= & \tilde{O}(m(m / r)^{1/3} r^{1/2} \epsilon^{-31/6})  \nonumber \\
=  & \tilde{O}(m^{4/3} r^{1/6} \epsilon^{-31/6})
\end{align*}

As $r \geq 1$, this cannot lead to an algorithm that runs faster
than $\tilde{O}(m^{4/3})$.
In order to obtain speedups, we need to reduce the $m$ term
in the computation of grouped flows.
This term corresponds to the number of edges in the linear
systems solved in each of the $\tilde{O}(k^{1/3}\epsilon^{-8/3})$
iterations needed for grouped flow computation.
We do so using the property that each group's boundary is
smaller than its size by a factor of about $r^{1/2}$.
For each subgraph $\component{i}$, we construct a smaller `equivalent'
of it that only contains vertices of $\VboundaryG{\component{i}}$.

The Schur complement described in Definition \ref{dfn:schurcomplement}
is a natural way to find such a smaller graph on boundary vertices.
However, the exact Schur complement is usually a dense graph 
with $(r^{1/2})^2 = r$ edges.
We obtain further reductions by showing that for approximating
grouped flows, a graph that's spectrally close to the Schur complement suffices.
This allows us to use the tool of spectral sparsification \cite{SpielmanS08},
which generates sparse approximations to such dense graphs.
Formally, we denote such approximations to the Schur complements
spectral vertex sparsifiers:

\begin{Definition}
\label{dfn:spectralvertexsparsifier}
Given a graph $G$ and a set of boundary vertices $\Vboundary(G)$.
A graph $\Gschurapprox$ on vertices $\VboundaryG{G}$ with corresponding Laplacian
$\laplacianpivotedapprox$ is an $\epsilon$-spectral vertex sparsifier for $(G, \VboundaryG{G})$
if:

\begin{enumerate}

\item $\Gschurapprox$ has $\tilde{O}(|\VboundaryG{G}|\epsilon^{-2})$ edges.

\item $\Gschurapprox$ and $\textsc{SchurComplement}(\laplacian, \Vinterior)$
are spectrally close.
\begin{align*}
(1 - \epsilon) \laplacianpivoted
\preceq &\laplacianpivotedapprox
\preceq (1 + \epsilon) \laplacianpivoted
\end{align*}

\item $\maxratio(\weightv(\Gschurapprox)) \leq \poly{n} \maxratio(\weightv(G))$.

\end{enumerate}

\end{Definition}

Our speedup is obtained by solving the grouped $L_2$ flow problem
on a graph with all groups replaced by their \textbf{spectral vertex sparsifiers}.
The conversion of flows to and from the smaller
graph is analyzed in Section \ref{sec:conversion}.
It leads an algorithm \textsc{ApproxGroupedFlow} whose running time
is about $\tilde{O}(m^{4/3} r^{-5/6} + m)$.
This can be viewed as depending inversely on $r$, the size of the groups,
or being proportional to the number of groups.

\begin{theorem}
\label{thm:approxgroupedflow}
Let $G = (V, E, \weightv)$ be a graph where there exists a grouped $L_2$
flow meeting demand $\demandv$ with maximum congestion $1 - \epsilon$.
Given a $r$-division of $G$ and $\epsilon/10$-spectral vertex sparsifiers for
all the groups, \textsc{ApproxGroupedFlow} returns
in $\tilde{O}(m^{4/3} r^{-5/6} \epsilon^{-14/3} \log(\maxratio(\weightv) / \epsilon) + m \log(\maxratio(\weightv) / \epsilon))$
time a flow meeting demands $\demandv$
where each group has congestion at most $1 + \epsilon$.
\end{theorem}

Note that when $r > O(m^{2/5})$, this total running time
is about $\tilde{O}(m)$.
Therefore, running time bottleneck now moves to the cost of
computing spectral vertex sparsifiers of the $O(m / r)$ groups,
each with $O(r)$ edges and $O(r^{1/2})$ boundary vertices.
In the rest of this section, We outline
our spectral vertex sparsification algorithms.
They're consequences of known results of using high
quality approximate solvers instead of exact inverses,
applied both once or recursively within the generalized
nested dissection framework.
Details and proofs on them are given
in Appendix \ref{sec:vertexsparsify}.

A direct approach is to compute the Schur complement as
Definition \ref{dfn:schurcomplement}, with sparsification done both
before and after computing the Schur complement.
Since $\laplacianinner$ is diagonally dominant,
we can invoke SDD linear system solvers
on $\laplacianinner$ with on each column of $\laplacianmiddle$.
This gives an almost-exact approximation to $\laplacianinner^{-1} \laplacianmiddle$
in about $\tilde{O}(r^{1.5})$ time.
Multiplying $\laplacianmiddle^T$ against this
$\tilde{O}(r^{1/2}) \times \tilde{O}(r^{1/2})$ matrix
takes $\tilde{O}(r^{3/2})$ time as
$\laplacianmiddle$ has $\tilde{O}(r)$ non-zero entries.
Analyzing the error incurred from the solver and applying spectral sparsification
at the end leads to a one-step construction of the spectral vertex sparsifier:

\begin{lemma}
\label{lem:onestep}
Given a graph $G = (V, E, \weightv)$ with $n$ vertices and $m$ edges
and error bound $0 < \epsilon < 1/2$.
There is a routine $\textsc{OneStepVertexSparsify}$ that
returns in $\tilde{O}(m + |\VboundaryG{G}|n \epsilon^{-2} \log(\maxratio(\weightv)))$ time
an $\epsilon$-spectral vertex sparsifier for $(G, \VboundaryG{G})$, $\Gschurapprox$,
such that $\maxratio(\weightv(\Gschurapprox)) \leq O(m^5 \maxratio(\weightv))$.
\end{lemma}

Combining this with Theorems \ref{thm:approxgroupedflow} and
\ref{thm:groupedflow} gives a total running time
of $\tilde{O}(mr\epsilon^{-9/2} + m^{4/3} r^{-1/3} \epsilon^{-43/6})$.
This is minimized when $r = m^{1/4} \epsilon^{-2}$, which in turn
gives a total of $\tilde{O}(m^{5/4} \epsilon^{-13/2})$.

This algorithm is applicable when we can find, or are given a good partition
of our initial graph, but cannot do so repeatedly in the rest of the graph.
For example, for shallow minor-free graphs (which are not guaranteed to
have recursive separator structures), the result of \cite{Wulffnilsen11}
allows us to find the top level $r$-division in $\tilde{O}(n^{5/4})$ time,
leading to an approximate maximum flow algorithm that runs in
$\tilde{O}(m^{5/4} \epsilon^{-6})$ time for shallow minor free graphs.
In the more general case where each group of the $r$-division has a larger
boundary of size $r^{\beta}$, this type of runtime savings still hold.
However, the gains stop at $\beta = 2/3$ which corresponds to 3D cubes.

For graphs which are recursively separable, we can take advantage of
the existence of a full separator tree to speed up the computation of the
spectral vertex sparsifier.
Specifically we an algorithm analogous to parallel nested dissection first
introduced by Pan and Reif \cite{PanR93}.
This approach ensures that every time we invoke the one-step algorithm
given in Lemma \ref{lem:onestep}, the boundary and interior are of
comparable size.
Interleaving sparsification at all stages with higher accuracy requirements
leads to an algorithm that runs in $\tilde{O}(r)$ time.

\begin{lemma}
\label{lem:recursive}
Let $G = (V, E, \weightv) $ be a graph with $n$ vertices and $m$ edges
given along with a $9/10$-separator tree $\mathcal{S}$.
Given a partition of the vertices $V$ into boundary and interior
vertices such that $|\VboundaryG{G}| \leq O(n^{1/2})$,
and an error parameter $\epsilon$,
$\textsc{VertexSparsify}(\laplacian, \VboundaryG{G}, \epsilon /  2\log_{20/19}{n})$
returns in $\tilde{O}(m + n \epsilon^{-2} \log{\maxratio(\weightv)})$
time an $\epsilon$-spectral vertex sparsifier for $(G, \VboundaryG{G})$, $\Gschurapprox$,
such that $\maxratio(\weightv(\Gschurapprox)) \leq O(m^5 \maxratio(\weightv))$.
\end{lemma}

This leads to a faster algorithm that finds an
$1 - \epsilon$ approximate maximum flow in
$\tilde{O}(m r^{1/2} \epsilon^{-9/2} + m^{4/3}  r^{-1/3} \epsilon^{-43/6})$
time on well-separated graphs.
This is minimized when $r = m^{2/5} \epsilon^{-16/5}$, giving a total running
time of $\tilde{O}(m^{6/5}\epsilon^{-61/10})$ and in turn Theorem \ref{thm:main}.

In order to give an overall picture of our algorithm, it's helpful to consider the case
where we're trying to route flows from the top left to bottom right on
a $n^{1/2} \times n^{1/2}$ square mesh.
The grid structure is chosen for this example in order to simplify
the description of the r-division.
Of course, in this case the faster planar maximum flow
algorithms can be used instead.
However, all the steps are essentially the same for separable graphs.

We use this example to walk through the main steps of our
algorithm on well-separated graphs.
Given the input graph shown in Figure~\ref{fig:mesh}  (a),
we decompose it into graphs with $r = O(n^{2/5})$ vertices,
of which $O(r^{1/2}) = O(n^{1/5})$ are on the boundary,
(in this case, smaller squares with side length $O(n^{1/5})$).
This leads to $k = n / r = O(n^{3/5})$ groups, giving the graph shown in (b).
To find an approximate grouped $L_2$ flow, we construct spectral vertex
sparsifiers using Lemma \ref{lem:onestep} or \ref{lem:recursive}.
These graphs have both fewer edges and vertices, giving (d).
An exact way of doing this sparsification  is to remove all internal nodes of each group
using partial Gaussian elimination, giving (c) where each group only has
the $O(r^{1/2}) = O(n^{1/5})$ boundary vertices.
However, these groups are now dense graphs and can have
$O((r^{1/2})^2) = O(r) = O(n^{2/5})$ edges per group.
However, the combination of this and spectral sparsification \cite{SpielmanS08}
gives $\tilde{O}(r^{1/2}) = \tilde{O}(n^{1/5})$ vertices and edges per group.
From the perspective of grouped $L_2$ flow problem
(b), (c), and (d) are roughly equivalent
except that (d) is substantially smaller.
Our algorithm then solves an approximate maximum flow problem on
the grid (a) by averaging a sequence of grouped flows.
The guarantees given in Lemma \ref{lem:groupedfloworacle}
leads to about $\tilde{O}(r^{1/2}) = \tilde{O}(n^{1/5})$
grouped $L_2$ flows on graph (b).
Each such instance is solved using the algorithm \textsc{ApproxGroupedFlow}
described in Theorem \ref{thm:approxgroupedflow}.
For each grouped $L_2$ flows on graph (b) we construct the smaller but
equivalent instance (d) as described above in $\tilde{O}(n)$ time.
We then solve the grouped $L_2$ problems on (d) using Theorem \ref{thm:groupedflow},
leading to a total running time of about $\tilde{O}(n^{4/3} r^{-5/6} + n) = \tilde{O}(n)$.
This running time can be further broken down into about
$\tilde{O}(k^{1/3}) = \tilde{O}(n^{1/5})$ electrical flow computations described
in Theorem \ref{thm:approxelectrical}.
These take place on (d), which has size about $O(n^{4/5})$.
Once we have this approximate grouped flow,
we can use the demands on the boundary vertices of each group
to convert it back to a flow in that group in (b) that meets the same demand.
Piecing these flows on all groups together gives a grouped flow in (b).
The total iteration count of our algorithm is
$\tilde{O}(r^{1/2}) \times \tilde{O}((n/r)^{1/3}) = \tilde{O}(n^{2/5})$.
Although this is slightly higher than the Christiano et al. algorithm,
our overall running time is faster due to the smaller size of (d).
Specifically, since $\tilde{O}(n^{1/5})$ calls to (b) are made, the total
running time becomes $\tilde{O}(n^{6/5})$.

\section{Using Approximate Grouped Flows}
\label{sec:approxgrouped}

Our algorithm uses approximate grouped flows in a way
similar to the use of electrical flows in the Christiano et al.
algorithm \cite{ChristianoKMST10}.
They showed that electrical flows can be used as an
$(\epsilon, \rho)$ oracle as defined below.

\begin{Definition}
\label{dfn:epsrho}
For $\epsilon > 0$ and $\rho > 0$, an $(\epsilon, \rho)$-flow oracle is an algorithm that
given demand $\demandv$, a flow amount $F$ and a vector $\weightoraclev$ of edge weights with
$\weightoracle(e) > 1$ for all $e$, returns one of the following:
\begin{enumerate}
\item If there exist a flow $\optflowv$ that routes $F$ units of flow from $s$ to $t$
and has max congestion $1$, then output a flow $\flowv$ satisfying:
\begin{enumerate}
\item $\flowv$ routes $F$ meets the demand, $\edgevertex^T \flowv = \demandv$
\label{cond:totalflow}
\item $\sum_e \weightoracle(e) \congestion{\flowv}{e} \leq (1 + \epsilon) |\weightoraclev|_1$,
where $|\weightoraclev|_1 = \sum_e \weightoracle(e)$;
\label{cond:totalcong}
\item $\congestion{\flow}{e} \leq \rho$ for all edges $e$.
\label{cond:maxcong}
\end{enumerate}
\item Otherwise, it either outputs a flow $\flowv$ satisfying conditions \ref{cond:totalflow}, \ref{cond:totalcong}, \ref{cond:maxcong} or outputs \fail.
\end{enumerate}
\end{Definition}

Christiano et al. showed that obtaining $(\epsilon, \rho)$-flow oracles is sufficient
for approximate maximum flow algorithms \cite{ChristianoKMST10}.
We use Theorem 3.2 from their paper as our starting point:

\begin{lemma}
\label{lem:oracleusage}
(Theorem 3.2 from  \cite{ChristianoKMST10})
For any $0 < \epsilon < 1/2$ and $\rho > 0$, given an $(\epsilon, \rho)$-flow oracle
with running time $T(m, 1/\epsilon)$,
one can obtain an algorithm that computes a $(1 - O(\epsilon))$-approximate
maximum flow in a capacitated, undirected graph in time
$\tilde{O}(\rho \epsilon^{-2}\cdot T(m, 1/\epsilon))$.
Furthermore, all calls to the $(\epsilon, \rho)$ oracle have
$\maxratio(\weightv) \leq O(m/\epsilon)$.
\end{lemma}

It was also shown Section 2.2 of \cite{ChristianoKMST10} that for finding
an approximate maximum flow, it suffices to consider capacities with
$\maxratio(\capacityv) \leq O(m / \epsilon)$.
We show that grouped flows can be used to give
$(\epsilon, \rho)$-flow oracles.
This is done by choosing the edge weights, $\weightv$
based on the weights given to the oracle, $\weightoraclev$.
The following proof is directly based on the on in Section
3.2. of \cite{ChristianoKMST10}, which shows that electrical
flows can be used as $(\epsilon, \rho)$-flow oracles.

\Proofof{Lemma \ref{lem:groupedfloworacle}}
By Lemma \ref{lem:oracleusage}, it suffices to provide an
$(\epsilon, O(r^{1/2}\epsilon^{-1/2}))$-flow oracle to the
Christiano et al. algorithm.
Let graph that we're given be $G = (V, E)$,
with edge capacities $\capacityv$;
the demand of the flow be $\demandv$ and
weights on edges be $\weightoraclev$.
Let $\weightoracle(\component{i})$ be the total weight among edges in component $i$,
aka. $\weightoracle(\component{i}) = \sum_{e \in E(\component{i})} \weightoracle(e)$.
We set the weight of an edge $e$ in component $i$ to:
\begin{align*}
\weight(e) = &
    \frac{1 - \epsilon / 2}{\capacity(e)^2}
        \left(\frac{\weightoracle(e)}{\weightoracle(\component{i})}
		+ \frac{\epsilon}{4|E(\component{i})|} \right)
\end{align*}

Note that the assignments of edge weights have an extra $\capacity(e)^2$
term in the denominator due to
$\congestion{\flowv}{e} = \frac{\flow(e)}{\capacity(e)}$.
This allows us to write the energy emitted on edge $e$
in the following more convenient form:
\begin{align*}
\weight(e) \flow(e)^2
= & \frac{1 - \epsilon / 2 }{\capacity(e)^2}
\left( \frac{\weightoracle(e)}{\weightoracle(\component{i})}
    + \frac{\epsilon}{4 |E(\component{i})|} \right)
	\flow(e)^2 \nonumber \\
= & (1 - \epsilon / 2) \left( \frac{\weightoracle(e)}{\weightoracle(\component{i})}
    + \frac{\epsilon}{4 |E(\component{i})|} \right)
	\congestion{ \flowv }{e}^2
\end{align*}

We first show that the existence of a flow $\optflowv$ where
$\congestion{\optflowv}{e} \leq 1$ implies that
$\congestion{\optflowv}{i} \leq 1 - \epsilon / 10$:
\begin{align*}
\congestion{\optflowv}{i}^2
= &  \sum_{e \in E(\component{i})} \weight(e) \optflow(e)^2 \nonumber \\
= &  (1 - \epsilon / 2) \sum_{e \in E(\component{i})}
\left( \frac{\weightoracle(e)}{\weightoracle(\component{i})} + \frac{\epsilon}{4 |E(\component{i})|} \right) \congestion{\optflowv}{e}^2 \nonumber \\
\leq &  (1 - \epsilon / 2) \sum_{e \in \component{i}}
\left(\frac{\weightoracle(e)}{\weightoracle(\component{i})} + \frac{\epsilon}{4 |E(\component{i})|} \right) 1\nonumber\\
\leq & (1 - \epsilon / 2) (1 + \epsilon / 4)\nonumber \\
\leq & 1 - \epsilon / 5
\end{align*}
Taking square roots of both sides gives the bound.

Thus if $\textsc{ApproxGroupedFlow}$ is run with an error bound
of $\epsilon / 10$, it returns a flow $\flowv$ where
$\congestion{\flowv}{i} \leq 1 + \epsilon / 10$.
This condition is equivalent to:
\begin{align*}
(1 + \epsilon / 10)^2
\geq & \congestion{\flowv}{i}^2 \nonumber \\
= & \sum_{e \in E(\component{i})} \weight(e) \flowv(e)^2 \nonumber \\
\geq & \weight(e) \flowv(e)^2 \nonumber \\
= & (1 - \epsilon / 2) \sum_{e \in E(\component{i})}
    \left(\frac{\weightoracle(e)}{\weightoracle(\component{i})}
    + \frac{\epsilon}{4 |E(\component{i})|} \right) \congestion{\flowv}{e}^2
\end{align*}
By the assumption that $\epsilon < 1/2$, we have
$\frac{(1 + \epsilon / 10)^2}{ 1 - \epsilon / 2} \leq ( 1 + \epsilon) ^ 2$.
Taking this into account and rearranging gives:
\begin{align*}
(1 + \epsilon)^2 \nonumber
\geq & \sum_{e \in E(\component{i})}
    \left(\frac{\weightoracle(e)}{\weightoracle(\component{i})}
    + \frac{\epsilon}{4 |E(\component{i})|} \right) \congestion{\flowv}{e}^2
\end{align*}

We now show that this flow meets the requirements of
an $(\epsilon, O(r^{1/2}\epsilon^{-1/2}) )$-flow oracle,
namely the following two conditions:
\begin{enumerate}
\item For each edge $e \in E(\component{i})$, $\congestion{\flowv}{e} \leq
O(|E(\component{i})|^{1/2} \epsilon^{-1/2}) \leq O(r^{1/2} \epsilon^{-1/2})$
\label{item:maxcong}
\item $\sum_{e \in E(\component{i})} \weightoracle(e) \congestion{\flowv}{e}
\leq (1 + \epsilon) \weightoracle(\component{i})$.
\label{item:totalcong}
\end{enumerate}

For Part \ref{item:maxcong}, we take the first term in the weights
and isolate edge $e$ to get:
\begin{align*}
(1 + \epsilon)^2
\geq & \frac{\epsilon}{4 |E(\component{i})|} \congestion{\flowv}{e}^2
\end{align*}

Rearranging gives:
\begin{align*}
\congestion{\flowv}{e} ^2
\leq & (1 + \epsilon)^2 4 |E(\component{i})| \nonumber \\
\leq & O( \epsilon^{-1} |E(\component{i})|)
\end{align*}
Taking square-roots of both sides again completes Part 1.

For Part \ref{item:totalcong}, we take the second part of the weight
terms to get:
\begin{align*}
(1 + \epsilon)^2
\geq & \sum_{e \in E(\component{i})}
    \frac{\weightoracle(e)}{\weightoracle(\component{i})} \congestion{\flowv}{e}^2
\end{align*}

Multiplying both sides again by $\weightoracle(\component{i})^2 =
(\sum_{e \in E(\component{i})} \weightoracle(e))^2$
and applying the Cauchy-Schwarz inequality gives:
\begin{align*}
(1 + \epsilon)^2 \left(\sum_{e \in E(\component{i})} \weightoracle(e) \right)^2 \geq & \left( \sum_{e \in E(\component{i})} \weightoracle(e) \right)
\left( \sum_{e \in E(\component{i})} \weightoracle(e) \congestion{\flowv}{e}^2 \right) \nonumber \\
\geq & \left( \sum_{e \in E(\component{i})} \weightoracle(e)\congestion{\flowv}{e} \right)^2
\end{align*}
Taking square roots of both sides gives Part \ref{item:totalcong}.

It remains to bound the ratio between weights assigned to two
edges $e$ and $e'$, we have:
\begin{align*}
\weight(e')
\leq & \frac{1 - \epsilon / 2}{\capacity(e')^2} (1 + \epsilon)
\leq \frac{2}{\capacity(e')^2} \\
\weight(e)
\geq & \frac{1 - \epsilon/2}{\capacity(e)^2} \frac{\epsilon}{4|E(\component{i})|}
\geq \frac{\epsilon }{4m\capacity(e)^2}\\
\frac{\weight(e')}{\weight(e)}
\leq & \left(\frac{2}{\capacity(e')^2}\right) / \left(\frac{\capacity(e)^2 \epsilon}{4m}\right) \nonumber \\
\leq & \frac{8m}{\epsilon} \left( \frac{\capacity(e)}{\capacity(e')} \right)^2
\leq O(m^3 \epsilon^{-3})
\end{align*}
\QED

\section{Converting Flows to and from Sparsifier}
\label{sec:conversion}

We now show that an approximate grouped flow can
be obtained from an approximate grouped flow on
the graph where each group is replaced with
a spectral a vertex-sparsifier.
The differences between both vertices and edges of the
two graphs makes a direct, embedding based
mapping between flows on the edges difficult.
As a result, our conversion process relies on computing
locally optimal electrical flows and bounding their energy
dissipations using guarantees of spectral sparsifiers.
We only use the residue of the flows on the boundaries
of each piece on to make this conversion.
The energy of the optimum electrical flow given
in Lemma \ref{lem:optelectrical} then allows us to
(approximately) preserve the energy of flows between
the original graph and its spectral vertex sparsifier.

Recall that as goal is to find $s-t$ flows, we may assume
that both $s$ and $t$ are on boundaries.
More generally, we restrict ourselves to flows where
internal vertices have zero demand.
We first show that for the exact Schur complement, such flows
can be converted exactly.
The following lemma is an re-statement of results on bounding
the support of Steiner tree based support tree preconditioners
\cite{Gremban-thesis,MaggsMOPW05,KoutisMiller08}.
The usual proof is given in Appendix \ref{sec:schurproofs}
for completeness.

\begin{lemma}
\label{lem:pivotok}
For any demand vector $\demandv$ defined on the vertices of a graph
partitioned into boundary and inner vertices such that the only non-zero
demands are on the boundary vertices.
Let $\demandboundary$ be the restriction of $\demandv$ onto the boundary
vertices and $\laplacianpivoted$ be $\textsc{SchurComplement}(\laplacian, \Vinterior)$.
Then we have:
\begin{align*}
\demandv^T \laplacian^+ \demandv
= \demandboundary^T \laplacianpivoted^+ \demandboundary
\end{align*}
\end{lemma}

By Lemma \ref{lem:optelectrical}, these two terms correspond
to energies of electrical flows on the two graphs.
This shows that if all inner vertices have zero demand,
the demands on the boundary determine the minimum energy.
However,  due to the need to work with smaller problem instances,
we can only use an approximation of $\laplacianpivoted$.
Let this approximate spectral vertex sparsifier be
$\laplacianpivotedapprox$.
The approximation guarantee given by Definition
\ref{dfn:spectralvertexsparsifier} is 
$(1 - \epsilon) \laplacianpivoted
\preceq \laplacianpivotedapprox
\preceq (1 + \epsilon) \laplacianpivoted$.
This error of $1 \pm \epsilon$ can be
accounted for in the energy difference of the flows.
This leads to the following Lemma, which allows us
to convert between flows on two graphs whose
Schur complement are similar.
Note that since the spectral vertex sparsifier is only
on the boundary vertices, it's a Schur complement of itself.

\begin{lemma}
\label{lem:convertsingle}
Given a graph $G = (V(G), E(G), \weightv(G))$
and $H = (V(H), E(H), \weightv(H))$ be connected graphs,
with $\VboundaryG{G} = \VboundaryG{H}$ and:
\begin{align*}
\textsc{SchurComplement}(G, \VboundaryG{G})
\preceq & (1 + \epsilon) \textsc{SchurComplement}(H, \VboundaryG{H})
\end{align*}
If $\flowv(G)$ is a flow on $G$ that meets demand $\demandv(G)$
that's only non-zero on boundary vertices.
Then we can find $\flowv(H)$ on $H$ such that the residuals of
$\flowv(H)$ equals to $\demandboundaryv$ on boundary vertices,
and are $0$ in all interior vertices, and:
\begin{align*}
\eflow_{\flowv(H)}(\weightv(H))
\leq (1 + 3\epsilon) \eflow_{\flowv(G)}(\weightv(G))
\end{align*}
Furthermore, $\flowv(H)$ can be found in
$\tilde{O}(|E(G)| + |E(H)|\log(\maxratio(\weightv(H)) / \epsilon))$ time.
\end{lemma}

\Proof
Since $G$ and $H$ are connected, the null spaces of their Schur
complements onto the boundary are the same.
If we denote them as $\laplacianpivoted(G)$ and $\laplacianpivoted(H)$,
the giving condition becomes:
\begin{align*}
\laplacianpivoted(H) \preceq (1 + \epsilon) \laplacianpivoted(G)
\end{align*}
Let $\demandv(G)$ and $\demandv(H)$ be the vectors formed
by extending $\demandboundaryv$ on to $V(G)$ and $V(H)$.
Substituting $\demandboundaryv$ into this bound and
applying Lemma \ref{lem:pivotok} gives:
\begin{align*}
\demandv(H)^T \laplacian(H)^{+} \demandv(H)
= & \demandboundaryv^T \laplacianpivoted(H)^+ \demandboundaryv  \nonumber \\
\leq &  (1 + \epsilon) \demandboundaryv^T \laplacianpivoted(G)^+ \demandboundaryv \nonumber \\
= &  (1 + \epsilon) \demandv(G)^T \laplacian(G)^{+} \demandv(G)
\end{align*}

By Lemma \ref{lem:optelectrical}, $\eflow_{\flowv(G)}(\weightv(G))
\geq \demandv(G)^T \laplacian(G)^{+} \demandv(G)$.
This in turns implies that minimum energy flow that routes $\demandv(H)$
in $H$ has energy at most $(1 + \epsilon) \eflow_{\flowv(G)}(\weightv(G))$.
Invoking Theorem \ref{thm:approxelectrical} with $\delta = \epsilon$
then returns a flow whose energy is at most:
\begin{align*}
(1 + \epsilon)\eflow(\weightv(H))
\leq & (1 + \epsilon)^2\eflow(\weightv(G)) \nonumber \\
\leq & (1 + 3\epsilon) \eflow_{\flowv(G)}(\weightv(G))
\end{align*}
\QED

Note that since $\demandv(G) = \edgevertex(G) \flowv(G)$
and each edge belongs to one of the groups, it can be easily
decomposed into a set of $k$ demands, one per group.
We can find the residue from all edges in each group, and
apply this conversion lemma to them.
This leads to an algorithm for converting flows between
graphs where each group are spectrally equivalent.
Its pseudocode is shown in Algorithm \ref{algo:convert}.

\begin{algorithm}[ht]

\qquad

\textsc{ApproxGroupedFlow}
\vspace{0.05cm}

\underline{Input:}
Graphs $G$ and $H$, along with partitions of edges
into groups $\component{1}, \ldots, \component{k}$
and $\componenth{1}, \ldots, \componenth{k}$ such that
$\Vboundary(\component{i}) = \Vboundary(\componenth{i})$
and $\laplacianpivoted(\componenth{i})
\preceq (1 + \epsilon) \laplacianpivoted(\component{i})$.
Flow on $\flowv(G)$ meeting demand $\demandv(G)$ such that
$\demandv(G)$ is only non-zero on boundary vertices.
Error bound $\epsilon$.

\underline{Output:}
Flow on $H$, $\flowv(H)$ meeting $\demandv(H)$ such that
$\demandboundaryv(H) = \demandboundaryv(G)$,
$\demandv(H)$ is $0$ on all interior vertices,
and $\congestion{\flowv(G)}{i}
\leq (1 + 3\epsilon)\congestion{\flowv(H)}{i}$.

\vspace{0.2cm}

\begin{algorithmic}[1]
\STATE{Initialize $\flowv(H) = 0$}
\FOR{$i = 1 \ldots k$}
    \STATE{Compute the residuals of $\flowv(G)$ among edges in $E(\component{i})$, $\demandv^{(i)}(G)$}
	\STATE{Form $\demandv^{(i)}(H)$ by taking $\demandboundaryv^{(i)}(G)$ and
putting $0$ on all interior vertices}
	\STATE{$\flowv^{(i)}(H) \leftarrow \textsc{ElectricalFlow}(\component{i}, \demandv(i), \weightv(i))$}
\ENDFOR
\STATE{Sum $\flowv^{(1)}(H) \ldots \flowv^{(k)}(H)$ to form $\flowv(H)$}
\RETURN{$\flowv(H)$}
\end{algorithmic}

\caption{Algorithm for Converting flow between two graphs where
Schur complements of all groups are close}

\label{algo:convert}

\end{algorithm}

\begin{lemma}
\label{lem:convert}
Let $G$ and $H$ be graphs with edge partitions 
$\component{1}, \ldots, \component{k}$
and $\componenth{1}, \ldots, \componenth{k}$ such that
$\Vboundary{\component{i}} = \Vboundary{\componenth{i}}$
and $\laplacianpivoted(\componenth{i})
\preceq (1 + \epsilon) \laplacianpivoted(\component{i})$.
Given a flow $\flowv(G)$ meeting demand $\demandv(G)$
that's only non-zero on boundary vertices and error bound $\epsilon$,
$\textsc{Convert}(G, H, \flowv(G), \epsilon)$ produces in
$\tilde{O}(|E(G)| + |E(H)| \log(\maxratio(\weightv(H)) / \epsilon))$
time a flow $\flowv(H)$ such that:

\begin{itemize}
\item $\flowv(H)$ meeting $\demandv(H)$ such that
$\demandboundaryv(H) = \demandboundaryv(G)$,
$\demandv(H)$ is $0$ on all interior vertices.
\item $\congestion{\flowv(G)}{i}
\leq (1 + 3\epsilon)\congestion{\flowv(H)}{i}$.
\end{itemize}

\end{lemma}

\Proof
By the given spectral condition between
$\component{i}$ and $\componenth{i}$
and Lemma \ref{lem:convertsingle}, we have that:
\begin{align*}
\congestion{\flowv(H)}{i}^2
= & \eflow_{\flowv^{(i)}(H)}(\weightv(\componenth{i}) \nonumber \\
\leq & (1 + 3 \epsilon) \eflow_{\flowv^{(i)}(G)}(\weightv(\component{i})\nonumber\\
= & \congestion{\flowv(G)}{i}^2
\end{align*}
Taking square roots of both sides gives the bound
on congestion.

Note that since the computation of residuals is linear
and each edge belongs to one partition, we have:
\begin{align*}
\demandv(G)
= &\sum_{i = 1}^k \demandv^{(i)}(G) \nonumber \\
\demandv(H)
= &\sum_{i = 1}^k \demandv^{(i)}(H)
\end{align*}
Since $\demandboundaryv^{(i)}(G)  = \demandboundaryv^{(i)}(H)$
and $\demandv^{(i)}(H)$ are $0$ on interior vertices, we satisfy
the requirement on $\demandv(H)$.
\QED

The faster algorithm that takes advantage of the $\epsilon$-spectral
vertex sparsifiers is now clear.
We take the restriction of $\demandv$ onto the sparsified graph,
and find an approximate grouped $L_2$ flow on it.
Since each group now only consists of their boundary vertices,
they're smaller by a factor of $\tilde{O}(r^{1/2})$, which compensates
for the $\tilde{O}(k^{1/3})$ iteration count from Theorem
\ref{thm:groupedflow}.
Once an approximate grouped flow is found on the smaller graph,
we convert it to one that approximately meets
the constraints on the original, unsparsified graph.
Pseudocode of this algorithm is shown in Algorithm
\ref{algo:approxgroupedflow}.

\begin{algorithm}[ht]

\qquad

\textsc{ApproxGroupedFlow}
\vspace{0.05cm}

\underline{Input:}
Weighted graph $G=(V,E, \weightv)$ with an $r$-division
partitioning it into groups $\component{1}, \ldots, \component{k}$.
Demand vector $\demandv$ such that only vertices on boundary of groups
can have non-zero demand.
Error bound $\epsilon$.
$\epsilon / 10$ spectral vertex sparsifiers
$\laplacianpivotedapprox(\component{i})$ for all the groups.

\underline{Output:}
Approximate grouped $L_2$ flow on $G$ with congestion $1 + \epsilon$

\vspace{0.2cm}

\begin{algorithmic}[1]
\STATE{Construct $\Gschurapprox$ by combining $\laplacianpivotedapprox(\component{i})$ for $1 \leq i \leq k$.}
\STATE{Let $\demandschurv$ be the restriction of $\demandv$ onto the $V(\Gschurapprox)$}
\STATE{$\approxflowv \leftarrow \textsc{GroupedFlow}(\Gschurapprox, E(\component{1}) \ldots E(\component{k}), \demandv, \epsilon/2)$}
\STATE{Let $\flowv$ be $\textsc{Convert}(G, \Gschurapprox, \approxflowv, \epsilon/10)$}
\RETURN{$\flowv$}
\end{algorithmic}

\caption{Faster Approximate Grouped $L_2$ Flows Using $r$-divisions}

\label{algo:approxgroupedflow}

\end{algorithm}

\Proofof{Theorem \ref{thm:approxgroupedflow}}
We first show that the flow produced meets the requirement.
By the requirement of spectral vertex sparsifiers given in Definition
\ref{dfn:spectralvertexsparsifier}, we have:
\begin{align*}
(1 - \epsilon/10) \laplacianpivoted(G^{(i)})
\preceq & \Gschurapprox^{(i)}
\preceq (1 + \epsilon/10) \laplacianpivoted(G^{(i)})
\end{align*}
Assume there is a grouped $L_2$ flow in $G$, $\optflowv$ with
congestion at most $1 - \epsilon$.
Then applying Lemma \ref{lem:convert} with error $\epsilon/10$
shows the existence a grouped $L_2$ flow in $\Gschurapprox$
with congestion at most $( 1 - \epsilon) (1 + 3/10 \epsilon) \leq 1 - \epsilon/2$.
Therefore by the guarantees of the grouped flow algorithm given
in Theorem \ref{thm:groupedflow}, we get that $\approxflowv$
has congestion at most $1 + \epsilon/2$.
Applying Lemma \ref{lem:convert} in the reverse direction, this
time algorithmically, gives a grouped $L_2$ flow in $G$ with
congestion at most $1 + \epsilon$.

It remains to bound the running time, which consists of running \textsc{GroupedFlow}
on the vertex-sparsified graph, and converting the flow back.
The cost of conversion is bounded by $\tilde{O}(m\log(U \epsilon^{-1}))$ and
the cost of grouped $L_2$ flow follows by the total number of edges in the components
and $k = O(m / r)$.
\QED

\section{Comments / Future Work}

\label{sec:comments}

For readers who are familiar with the Christiano et al. algorithm,
an alternate view of our algorithm that's closer to the multiplicative weights
update framework is that it takes advantages of the grouping
by making more aggressive readjustments of all edges in the same group.
In this view we track and adjust another set of weights
defined for each component which are adjusted by the grouped $L_2$
flow algorithm, while the per edge weights are adjusted less frequently
in an outer layer.
In the electrical flow computations, the resistive value of an edge
then depends on both its weight and that of its component.

To better understand the costs of our algorithm it's helpful to give
a break down of the costs incurred by each component,
ignoring constants and logarithmic terms.
Such a list is given in Figure \ref{fig:summary}, where cost is
the total cost incurred by all calls made by the component and
number of calls is the total number of times this component is called
in order to compute an approximate maximum flow.
We now go over these procedures, their costs, and calling order.
To compute a maximum flow we average $O(r^{1/2})$,
the width of the oracle, grouped flows on the full graph.
The grouped flow procedure makes several calls.
It first computes spectral vertex sparsifier on each of the $ (n/r)$ groups.
Then it computes a grouped flow on the sparsified graph by calling
electrical flow $(n/r)^{1/3}$ times.
The grouped $L_2$ flow on the sparsified graph is then converted
to the full graph group-by-group.

\begin{figure}[ht]
\begin{center}
\begin{tabular}{|p{8cm}|p{1cm}|p{3cm}|p{1cm}|p{2cm}|}
Procedures  & Size & Number of Calls & Cost & Total Cost \\
\hline
\hline
Grouped $L_2$ Flow on full graph & $n$ & $r^{\frac{1}{2}}$ & n & $nr^{\frac{1}{2}}$ \\
\hline
Spectral Vertex Sparsification & $r$ & $r^{\frac{1}{2}}\times n/r$ & $r$ & $nr^{\frac{1}{2}}$ \\
\hline
Grouped $L_2$ Flow on Sparsifier & $nr^{-\frac{1}{2}}$ & $r^{\frac{1}{2}}$ & n & $nr^{\frac{1}{2}}$  \\
\hline
Electrical Flow on Sparsifier & $ nr^{-\frac{1}{2}}$ & $r^{\frac{1}{2}} \times (n/r)^{\frac{1}{3}}$ &  $nr^{-\frac{1}{2}}$& $ n^{\frac{4}{3}}r^{-\frac{1}{3}}$ \\
\hline
Extrapolation of grouped $L_2$ flow to full graph & $r$ & $r^{\frac{1}{2}}\times \frac{n}{r}$ & $r$ & $nr^{\frac{1}{2}}$ \\
\end{tabular}
\caption{Summary of components and their costs in a sparse graph parameterized
in terms of $r$, the size of groups in the $r$-division}
\label{fig:summary}
\end{center}
\end{figure}

Our recusive approach extends readily to multiple levels,
but  does not lead to better running times when we
make more gradual reductions in graph sizes.
The main reason is that the computation of spectral vertex sparsifiers
in Appendix \ref{sec:vertexsparsify} first computes a dense and
almost exact vertex sparsifier.
The size of this dense graph at $O(|\VboundaryG{\component{i}}|^2)$ is the
bottleneck, although our algorithm only uses a sparsified version of it.
We believe constructing the final spectral vertex sparsifier
directly from the original graph is an interesting question on its own.
On the other hand, for graphs with better separator structures
such as graphs with bounded or low tree-width, this term is much
smaller and improved running times are likely.

Another possible way to obtain a speedup is to reduce the width of
grouped flows to $O(r^{1/3} \poly{1/\epsilon})$ in a way similar to the
Christiano et al. algorithm \cite{ChristianoKMST10}.
The difficulty here is that their analysis requires tracking terms up
to precision of roughly $r^{-1/3}$, but going through spectral
vertex sparsifiers incurs an error or $\epsilon$ already.
As a result, improvements in this direction will likely require a more
robust analysis of the Christiano et al. algorithm.

Finally, it's worth noting that all known hard instances
for the current analysis of the Christiano et al. algorithm,
such as the one given in Section 4 of \cite{ChristianoKMST10}
have good separator structures.
Whether this type of divide and conquer approach can lead
to speed-ups for approximate maximum flow on general graphs
is an intriguing direction for future work.

\section*{Acknowledgements}

We thank the anonymous reviewers for their very helpful comments.




\begin{spacing}{0.7}
  \begin{small}
    \bibliographystyle{alpha}
    \newcommand{\etalchar}[1]{$^{#1}$}

  \end{small}
\end{spacing}

\begin{appendix}
\section{Algorithm for Approximating Grouped $L_2$ Flow}
\label{sec:groupedflow}

We now show that given an undirected flow instance where
the edges are grouped into $k$ groups,
the maximum $L_2$ energy among the groups can be approximately
minimized using about $\tilde{O}(k^{1/3}\epsilon^{-8/3})$ iterations.
The algorithm assigns weights to all the groups and
adjusts them in an iteratively.
We will use $^{(t)}$ to denote that
the variable is associated with the $t$-th iteration.
Pseudocode of the algorithm is shown in Algorithm \ref{alg:groupedflow}.

\begin{algorithm}[ht]

\textsc{GroupedFlow}
\vspace{0.05cm}

\underline{Input:}
Weighted, undirected graph $G=(V,E,S_1, \ldots, S_k)$
with edges partitioned into $S_1 \ldots S_k$ and
weights on edges given by $\weightv: E \rightarrow \Re^+$.
Demand vector $\demandv$ and error bound $\epsilon$.

\underline{Output:} Either a flow $\flowv$ that meets the demands given
by $\demandv$ and that $\congestion{\flowv}{i} \leq 1 + 10 \epsilon$.
Or \fail~indicating that no such flow with maximum congestion $1$ exists.

\vspace{0.2cm}

\begin{algorithmic}[1]

\STATE{$\rho \leftarrow 10 k^{1/3} \epsilon^{-2/3}$}
\STATE{$\numiter \leftarrow 20 \rho \ln{(k)} \epsilon^{-2} = 200 k^{1/3} \ln{(k)} \epsilon^{-8/3}$}
\STATE{Initialize $\weightgroup^{(0)}(i) = 1$ for all groups $1 \leq i \leq k$}
\STATE{$\flowv \leftarrow \zerosv$, $\numiter_1 \leftarrow 0$}
\FOR{$t = 1 \ldots \numiter$}
	\STATE{Compute $\mu^{(t-1)} \leftarrow \sum_i \weightgroup^{(t-1)}(i)$} \label{ln:defmu}	
      \STATE{Set $\resistance(e)$ to $(\weightgroup^{(t-1)}(i) + \frac{\epsilon}{k} \mu^{(t - 1)}) \cdot \weight(e)$
		for all edges $e$ in group $S_i$}
	\STATE{$\approxflowv^{(t)} = \textsc{ElectricalFlow}(G=(V, E), \demandv, \resistancev)$}\label{ln:delta1}
	\IF{$\eflow_{\approxflowv^{(t)}}(\resistancev) > \mu^{(t-1)}$}
		\RETURN{\fail}
	\ELSE
	\IF{$\congestion{\flowv}{i} \leq \rho$ for all $i$}
		\STATE{$\flowv \leftarrow \flowv + \approxflowv^{(t)}$}
		\STATE{$\numiter_1 \leftarrow \numiter_1 + 1$}
       \ENDIF
		 \STATE{Update weights for all groups,
			$\weightgroup^{(t)}(i) \leftarrow \weightgroup^{(t-1)}(i)
			\left(1+\frac{\epsilon}{\rho} \congestion{\flowv}{i}\right) $}
	\ENDIF
\ENDFOR
\RETURN {$\frac{1}{\numiter_1} \flowv$}

\end{algorithmic}

\caption{Algorithm for grouped flows}

\label{alg:groupedflow}
\end{algorithm}

We first state the following bounds regarding the overall sum of potentials $\mu^{(t)}$,
the weight of a single group  $\weight^{(t)}(i)$ and the effective conductance given by
the reweighed energy matrices at each iteration, $\epotential^{(t)}$.

\begin{lemma}
\label{lem:potentials}
The following holds when $\approxflowv^{(t)}$ satisfies
\begin{align*}
\sum_i \weightgroup^{(t)}(i) \congestion{\approxflowv^{(t)}}{i} \leq \mu^{(t-1)}
\end{align*}

\begin{enumerate}
    \item \label{part:muupper}
        $\mu^{(t)} \leq \exp \left( \frac{\epsilon}{\rho} \right) \mu^{(t-1)}$
    \item \label{part:weightlower}
        $\weightgroup^{(t)}(i)$ is non-decreasing across iterations $t$, and if
$\congestion{\approxflowv^{(t)}}{i} \leq \rho$
		we have:
            \begin{align*}
                 \weightgroup^{(t)}(i) \geq \exp \left(\frac{\epsilon}{\rho} 
				\congestion{\approxflowv^{(t)}}{i} \right) \weightgroup^{(t-1)}(i)
            \end{align*}
    \item \label{part:potentialincrease}
        If for some edge $e$ we have $\congestion{\approxflowv^{(t)}}{i} \geq \rho$, then
            $\epotential(\resistancev^{(t)}) \geq \epotential(\resistancev^{(t-1)}) \exp \left( \frac{\epsilon^2 \rho^2}{4k} \right)$
\end{enumerate}
\end{lemma}

The proof of Lemma \ref{lem:potentials} relies on the following facts about $\exp(x)$ when $x$ is close to 1:

\begin{fact}
\label{fact:log}
\begin{enumerate}
   \item \label{part:logupper}
   If $x \geq 0$, $1 + x \leq \exp(x)$.
   \item \label{part:loglower}
   If $0 \leq x \leq \epsilon$, then $1 + x \geq \exp((1-\epsilon)x)$.
\end{enumerate}
\end{fact}

\Proofof{Part \ref{part:muupper}}
\begin{align*}
\mu^{(t)}
= & \sum_i \weightgroup^{(t)}(i) \nonumber \\
= & \sum_i \weightgroup^{(t-1)}(i)
	(1+\frac{\epsilon}{\rho} \congestion{\approxflowv^{(t)}}{i})
   \qquad \text{By the update rule} \nonumber \\
= & \left( \sum_i \weightgroup^{(t-1)}(i) \right)
    + \frac{\epsilon}{\rho} \sum_i \weightgroup^{(t-1)}(i) \congestion{\approxflowv^{(t)}}{i} \nonumber \\
 = & \mu^{(t - 1)}
    + \frac{\epsilon}{\rho} \sum_i \weightgroup^{(t-1)}(i) \congestion{\approxflowv^{(t)}}{i} 
    \qquad \text{By definition of $\mu^{(t-1)}$} \nonumber \\
\leq & \mu^{(t-1)} + \frac{\epsilon}{\rho} \mu^{(t-1)}
    \qquad \text{By bound on total weighted congestion} \nonumber \\
= & (1 + \frac{\epsilon}{\rho}) \mu^{(t-1)} \nonumber \\
\leq & \exp(\frac{\epsilon}{\rho}) \mu^{(t-1)}
  \qquad \text {By Fact \ref{fact:log}.\ref{part:logupper}}
\end{align*}
\QEDpart{Part \ref{part:muupper}}

\Proofof{Part \ref{part:weightlower}}

If $\congestion{\approxflowv^{(t)}}{i} \leq \rho$, then $\frac{\epsilon}{\rho} \congestion{\approxflowv^{(t)}}{i} \leq \epsilon$
and:
\begin{align*}
\weightgroup^{(t)}(i)
= & \weightgroup^{(t-1)}(i) \left( 1 + \frac{\epsilon}{\rho} \congestion{\approxflowv^{(t)}}{i} \right) \nonumber \\
\leq & \weightgroup^{(t-1)}(i) \exp \left( \frac{\epsilon (1-\epsilon) }{\rho} \congestion{\approxflowv}{i} \right)
  \qquad \mbox{By Fact \ref{fact:log}.\ref{part:loglower}}
\end{align*}
\QEDpart{Part \ref{part:weightlower}}

In order to prove Part \ref{part:potentialincrease}, we need the following
lemma was proven in \cite{ChinMMP12} about increasing the weight of
a heavily congested edge.

\begin{lemma}
\label{lem:dualincrease}
Let $\optflowv$ be the optimal electrical flow with flow value $F$
on a graph $G$ with resistances $\resistancev$.
Suppose there is a subset of the edges $S \subseteq E$ that accounts for $\beta$
of the total energy of $\optflowv$, i.e.
\begin{align*}
\sum_{e \in S} \optflow(e)^2 \resistance(e) \geq \beta \eflow(\resistancev, \optflowv)
\end{align*}
For some $\gamma > 0$, define new resistances $\resistancev'$ such that
$\resistance'(e) = (1+\epsilon) \resistance(e)$ for all $e \in S$
and $\resistance'(e) = \resistance(e)$ for all $e \notin S$, then
\begin{align*}
\eflow(\resistancev') \leq \exp(-\frac{\epsilon \beta}{2}) \eflow(\resistancev)
\end{align*}
\end{lemma}

\Proofof{Part \ref{part:potentialincrease}}

Let $i$ be the group such that $\congestion{\approxflowv}{i} \geq \rho$, then
since $\weightgroup(i) \geq \frac{\epsilon}{k} \mu$:
\begin{align*}
\weightgroup^{(t - 1)}(i) \congestion{\approxflowv^{(t)}}{i}^2
\geq & \frac{\epsilon}{k} \mu^{(t-1)} \rho^2 \nonumber \\
\geq & \frac{\epsilon \rho^2}{k} \eflow(\resistancev^{(t-1)}, \approxflowv)
\qquad \mbox{By assumption of the energy of the flow returned} \label{eq:energyfraction}
\end{align*}

Invoking the guarantees given in proven in Part \ref{part:flowdifference}
of Theorem \ref{thm:approxelectrical}, we have:
\begin{align*}
\sum_{e \in S_i} \resistance^{(t)}(e) \optflow(e)^2
\geq & \sum_{e \in S_i} \resistance^{(t - 1)}(e) \congestion{\approxflowv}{e}^2
    -    \left| \resistance^{(t - 1)}(e)\congestion{\optflowv}{e}^2
		- \resistance^{(t - 1)}(e)\congestion{\approxflowv}{e}^2 \right| \nonumber \\
\geq & \congestion{\approxflowv^{(t)}}{i} - \delta  \eflow(\resistancev^{(t-1)})
	\qquad \mbox{By Part \ref{part:flowdifference} of Theorem \ref{thm:approxelectrical}} \nonumber\\
\geq & \frac{\epsilon \rho^2}{k} \eflow_{\approxflowv^{(t)}}(\resistancev^{(t-1)})) -  \delta  \eflow(\resistancev^{(t-1)})
	\qquad \mbox{By Equation \ref{eq:energyfraction}} \nonumber \\
\geq & \frac{\epsilon \rho^2}{(1+\delta)k} \eflow(\resistancev^{(t-1)}) - \delta \eflow(\resistancev^{(t-1)})
	\qquad \mbox{By Part \ref{part:flowenergy}} \nonumber \\
\geq & \frac{\epsilon \rho^2}{2k} \eflow(\resistancev^{(t-1)})
\end{align*}

Applying Lemma \ref{lem:dualincrease} with $\beta =  \frac{\epsilon \rho^2}{4k}$
completes the proof.

\QEDpart{Part \ref{part:potentialincrease}}

\Proofof{Theorem \ref{thm:groupedflow}}

Since $\sum_{1 \leq i \leq k} (\weightgroup^{(t-1)}(i) + \frac{\epsilon}{k} \mu^{(t-1)})  = (1+ \epsilon)\mu^{(t-1)}$,
if there exist a flow $\flowv$ such that $\congestion{\flowv}{i} \leq 1 - 2 \epsilon$
for all $e$, we have that $\eflow(\resistancev^{(t)}) \leq (1 - \epsilon) \mu^{(t-1)}$.
Then if the algorithm does not return $\fail$, Theorem \ref{thm:approxelectrical}
means that $\approxflowv^{(t)}$ satisfies:
\begin{align*}
\eflow_{\approxflowv^{(t)}}(\resistancev^{(t)})
\leq & (1+\delta) (1 - \epsilon) \mu^{(t-1)} \nonumber \\
\leq & \mu^{(t-1)} \\
\sum_{i} \weightgroup^{(t-1)}(i) \congestion{\approxflowv^{(t)}}{i}^2
    \leq & \sum_{i} \weightgroup^{(t-1)}(i)
\end{align*}
Multiplying both sides by $\mu^{(t-1)}$ and
applying the Cauchy-Schwarz inequality gives:
\begin{align*}
\left( \sum_{i} \weightgroup^{(t-1)}(i) \right)^2
\geq & \left( \sum_{i} \weightgroup^{(t-1)}(i) \right)
	+\left( \sum_{i} \weightgroup^{(t-1)}(i) \congestion{\approxflowv^{(t)}}{i}^2 \right) \nonumber \\
\geq & \left( \sum_{i} \weightgroup^{(t-1)}(i) \congestion{\approxflowv^{(t)}}{i} \right)
\end{align*}
Taking the square root of both sides gives:
\begin{align*}
\sum_{i} \weightgroup^{(t-1)}(i) \congestion{\approxflowv^{(t)}}{i}
\leq & \mu^{(t-1)}
\end{align*}

Therefore inductively applying Lemma \ref{lem:potentials} Part\ref{part:muupper},
we have:
\begin{align*}
\mu^{(\numiter)}
\leq & \mu^{(0)} \cdot \left( \exp (\frac{\epsilon}{\rho}) \right)^\numiter \nonumber \\
= & \exp \left( \frac{\epsilon \numiter}{\rho}\right) m \nonumber \\
\leq & \exp \left( \frac{21 \ln m}{\epsilon} \right)
\end{align*}

We now bound $N'$, the number of iterations $t$ where there is a group with
$\congestion{\approxflowv^{(t)}}{i} \geq \rho$.
Suppose $\epotential(\resistancev^{(0)}) \leq 1/2$, then in the flow returned,
no group has $\congestion{\approxflowv^{(0)}}{i} \geq 1$,
which means that the algorithm can already return that flow.

Note that $\resistancev^{(t)}$ is monotonically increasing on each edge,
so $\epotential(\resistancev^{(t)})$ is also monotonic.
Then by Lemma \ref{lem:potentials} Part \ref{part:potentialincrease}, we have:
\begin{align*}
\epotential(\resistancev^{(\numiter)})
\geq & 1/2 \cdot \exp \left( \frac{\epsilon^2 \rho^2 \numiter'}{4m} \right)
\end{align*}
Combining this with
$\epotential(\resistancev^{(t)})
\leq \mu^{(\numiter)} \nonumber$ gives:
\begin{align*}
\frac{\epsilon^2 \rho^2 \numiter'}{4m}
\leq & \frac{21 \ln m}{\epsilon} \nonumber \\
\numiter'
\leq & \frac{100 m \ln m}{\rho^2  \epsilon^3}
\leq \epsilon {\numiter}
\end{align*}

Then in all the $\numiter - \numiter' \geq (1-\epsilon)\numiter$ iterations, we have
$\congestion{\approxflowv^{(t)}}{i} \leq \rho$ for all groups $i$.
This allows us to finish our proof by bounding the congestion of
each group:
\begin{align*}
\congestion{\sum_{t} \approxflowv^{(t)}}{i}
\leq & \sum_{t} \congestion{\approxflowv^{(t)}}{i}
\qquad \mbox{Since $\congestion{\cdot}{i}$ is a $L_2$ norm} \nonumber \\
\leq & \log(\mu^{(t)}) / ( \frac{\epsilon}{\rho} )
\qquad \mbox{By Lemma \ref{lem:potentials} Part \ref{part:weightlower}} \nonumber \\
= & \frac{1}{1 - \epsilon} T' \leq (1 + 2\epsilon) T'
\end{align*}
\QED

\section{Spectral Vertex Sparsification}
\label{sec:vertexsparsify}

This section described faster algorithms for spectral vertex sparsification.
As the Schur complements of the Laplacian of
a disconnected graph is the sum of the Schur complement
of its components' graph Laplacians,
we assume that the graphs given as input are connected.
We will use the following lemma about the Schur
complements of spectrally similar PSD matrices.

\begin{lemma}
\label{lem:approxschur}
Suppose $\laplacian(G)$ and $\laplacian(H)$ are matrices such that:
\begin{align*}
(1-\epsilon) \laplacian(G)
\preceq \laplacian(H)
\preceq (1+\epsilon) \laplacian(G)
\end{align*}

Then let $\laplacianpivoted(G)$ and $\laplacianpivoted(H)$ be the Schur
complements of  $\laplacian(G)$ and $\laplacian(H)$ on the same
set of vertices.
We have:
\begin{align*}
(1-\epsilon) \laplacianpivoted(G)
\preceq \laplacianpivoted(H)
\preceq (1+\epsilon) \laplacianpivoted(G)
\end{align*}
\end{lemma}

The following fact is crucial in relating energy of the Schur complement to that
of the original graph.
It can be viewed as an instance of the Dirichlet Principle (see \cite{doylesnell84},
page 64), which states that enforcing intermediate voltages other than the
naturally occurring ones only increases total power.
For completeness proof of it is included in Appendix \ref{sec:schurproofs}

\begin{lemma}
\label{lem:schurextended}
\begin{align*}
\vecx^T \laplacianpivoted(G) \vecx
=& \min_{\vecy}
\vecyx^T
\laplacian(G)
\vecyx
\end{align*}
\end{lemma}

\Proofof{Lemma \ref{lem:approxschur}}
We first show lower bound on $\laplacianpivoted(H)$.
For a fixed vector $\vecx$, let $\vecy$ the vector
given by Lemma \ref{lem:schurextended} for
$\laplacian(H)$.
Then we get:
\begin{align*}
(1-\epsilon) \vecx^T \laplacianpivoted(G) \vecx
\leq &
(1-\epsilon)
\vecyx^T
\laplacian(G)
\vecyx \\
\leq &
\vecyx^T
\laplacian(H)
\vecyx\\
= & \vecx^T \laplacianpivoted(H) \vecx
\end{align*}

For the upper bound on $\laplacianpivoted(H)$,
once again let $\vecy$ be given by Lemma
\ref{lem:schurextended} for $\laplacian(G)$ and $\vecx$.
This gives:
\begin{align*}
(1+\epsilon) \vecx^T \laplacianpivoted(G) \vecx
= &
(1+\epsilon)
\vecyx^T
\laplacian(G)
\vecyx \\
\geq &
\vecyx^T
\laplacian(H)
\vecyx\\
\geq & \vecx^T \laplacianpivoted(H) \vecx
\end{align*}
\QED

We can also establish bounds relating the spectrum of $\laplacianpivoted$
to the spectrum of $\laplacian$.

\begin{lemma}
\label{lem:schurspectrum}
Let $\laplacianpivoted = \textsc{SchurComplement}(\laplacian, \Vinterior)$,
then:
\begin{align*}
\lambda_2(\laplacianpivoted) \geq & \lambda_2(\laplacian)\\
\lambda_n(\laplacianpivoted) \leq & n\lambda_n(\laplacian)
\end{align*}

\end{lemma}

Proof of it is given in Appendix \ref{sec:schurproofs}.

Recall from Definition \ref{dfn:schurcomplement}
that the Schur complement can be obtained by partitioning
$\laplacian$ into blocks on the boundary and inner vertices:
\begin{align*}
\laplacian =
&\left [
\begin{array}{cc}
\laplacianinner & \laplacianmiddle \\
\laplacianmiddle^T & \laplacianboundary\\
\end{array}
\right]
\end{align*}
And evaluating $\laplacianpivoted
= \laplacianboundary - \laplacianmiddle^T \laplacianinner^{-1} \laplacianmiddle$.
Note that $\laplacianinner$ is a graph Laplacian on the inner vertices
with additional diagonal entries.
This places it within the class of symmetric diagonally dominant
(SDD) matrices.
Furthermore, due to the graph being non-connected it is strictly SDD.
We first show that the solves involving $\laplacianinner$
can be done using faster SDD linear system solvers.
We use the following result by Spielman and Teng on nearly-linear
time solvers for such systems:

\begin{lemma}
\label{lem:solver}
\cite{SpielmanTeng04, SpielmanTengSolver, KoutisMP_FOCS10, KoutisMP_FOCS11}.
Given a strictly SDD matrix of the form $\mata$,
and an error parameter $\delta$,
there is a symmetric linear operator $\matb$ such that
\begin{align*}
(1 - \delta) \mata^{-1} \preceq \matb \preceq (1+\delta) \mata^{-1}
\end{align*}
Furthermore, for any vector $\vecx$, $\matb \vecx$ can be evaluated in time
$\tilde{O}(m \log(1/\delta))$ where $m$ is the number of non-zero entries in $\mata$.
\end{lemma}

A direct approach is to apply the linear operator corresponding to
$\laplacianboundary^{-1}$ to $\laplacianmiddle$.
As this is done one column at a time, we use $\laplacianmiddle[:, i]$
to denote the $i$-th column of $\laplacianmiddle$.
Pseudocode for performing approximate Schur complements is given
in Algorithm \ref{alg:approxschur}

\begin{algorithm}[ht]
\qquad

\textsc{ApproxSchur}
\vspace{0.05cm}

\underline{Input:}
Graph $G = (V, E, \weightv)$ with corresponding Laplacian $\laplacian$,
boundary vertices $\Vboundary$.
Spectrum bound $\maxcond$ such that $\lambda_{n}(\laplacian)\
\leq \maxcond \lambda_{2}(\laplacian)$,
error parameter $\epsilon$.

\underline{Output:}
Approximate Schur complement $\laplacianpivotedapprox$.

\begin{algorithmic}[1]
\STATE{Let $\Vinterior$ denote the interior vertices, $V \setminus \Vboundary$}
\STATE{$\delta \leftarrow 2 \epsilon n^{-1} \kappa^{-1}$} \label{ln:delta} 
\FOR{$i = 1 \ldots |\Vboundary|$}
	\STATE{$\maty[:, i] \leftarrow \textsc{Solve}(\laplacianinner, \laplacianmiddle[:, i], \delta)$} 
\ENDFOR
\STATE{$\laplacianpivotedapprox'  \leftarrow \laplacianboundary  -
	 \laplacianmiddle^T \maty$}
\STATE{Let $\laplacianpivotedapprox$ be $\laplacianpivotedapprox'$ with positive
off-diagonal entries replaced by $0$s and diagonal entries readjusted to weighted degrees}
\RETURN{$\laplacianpivotedapprox$}
\end{algorithmic}

\caption{Approximate Schur Complement Using SDD Linear System Solvers}

\label{alg:approxschur}
\end{algorithm}

Note that since the linear operator corresponding to $\textsc{Solve}$
is symmetric, $\laplacianpivotedapprox'$ is symmetric.
However, $\textsc{Solve}$ only produces an approximation to the inverse.
We next show that setting $\delta$ to $\poly{n^{-1}, \maxratio(\weightv)^{-1}}$
suffices for approximating the Schur complement.
We start with rough bounds on the maximum weight
based on the eigenvalues.

\begin{lemma}
\label{lem:roughbound}
Let G be a graph with corresponding graph Laplacian $\laplacian$.
The maximum weight of an edge in $\laplacian$ is at most $2 \lambda_{n}$,
and for any choice of boundary vertices, $\laplacianboundary \preceq
2n \lambda_{n} \mati$.
\end{lemma}

\Proof
By the Courant-Fisher theorem applied to the vector with $1$ on one end and
$-1$ on the other, we get that its weight is at most $2\lambda_{n}$.

Since $\laplacianboundary$ is formed by taking a subset of the edges
and adding weighted degrees to the diagonal, we have:
\begin{align*}
\laplacianboundary
\preceq & \laplacian + n \lambda_{n} \mati
\preceq 2n \lambda_{n} \mati
\end{align*}
\QED

This allows us to prove the multiplicative error guarantee.

\begin{lemma}
\label{lem:approxschurcomplement}
Given a graph Laplacian $\laplacian$ for a graph $G$ with $n$ vertices
and $m$ edges, along with spectral bound $\maxcond$
such that $\lambda_{n}(\laplacian)\
\leq \maxcond \lambda_{2}(\laplacian)$,
and error bound $0 < \epsilon < 1/2$.
Let $\laplacianpivoted$ be the exact Schur complement of $(G, \Vinterior)$.
Then  $\textsc{ApproxSchur}(G, \Vboundary, \maxcond, \epsilon)$
in $\tilde{O}(|\Vboundary| m \log(\kappa \epsilon^{-1}) )$ time
outputs a matrix $\laplacianpivotedapprox$ such that:
\begin{align*}
(1 - \epsilon) \laplacianpivoted
\preceq \laplacianpivotedapprox \preceq
(1+\epsilon) \laplacianpivoted
\end{align*}
\end{lemma}

\Proof

We only show the LHS inequality as the RHS side follows similarly.
Let $\matb$ be the linear operator generated by $\textsc{Solve}$.
Then by the guarantees given in Lemma \ref{lem:solver} we have
$\matb \preceq (1 + \delta) \laplacianinner^{-1}$.
This in turn gives $\laplacianmiddle^T \matb \laplacianmiddle
\preceq (1 + \delta) \laplacianmiddle^T  \laplacianinner^{-1} \laplacianmiddle$
and:
\begin{align*}
\laplacianpivotedapprox
= & \laplacianboundary - \laplacianmiddle^T \matb \laplacianmiddle \nonumber \\
\succeq & \laplacianboundary - (1 + \delta) \laplacianmiddle^T 
\laplacianinner^{-1} \laplacianmiddle \nonumber \\
= & \laplacianpivoted - \delta \laplacianmiddle^T \laplacianinner^{-1} \laplacianmiddle
\end{align*}
Since $\laplacianboundary - \laplacianmiddle^T \laplacianinner^{-1} \laplacianmiddle$
is the Schur complement and is positive semi-definite,
we may use $\delta \laplacianboundary$ as an upper bound for
the difference.
Specifically we can use the bound on $\laplacianboundary$ given above.
Combining it with $\kappa_2(\laplacianpivoted) \leq \lambda_2(\laplacian)$
from Lemma \ref{lem:schurspectrum} finishes the bound:
\begin{align*}
\laplacianpivotedapprox
\succeq & \laplacianpivoted -  \delta 2n \lambda_{n}(\laplacian) \mati \nonumber\\
\succeq & \laplacianpivoted -  \epsilon \lambda_2(\laplacian) \mati \nonumber\\
\succeq & \laplacianpivoted -  \epsilon \lambda_2(\laplacianpivoted) \mati \nonumber\\
\succeq & (1 - \epsilon) \laplacianpivoted
\end{align*}

We now bound the running time.
The inner loop consists of $|\Vboundary|$ calls to \textsc{Solve},
each taking at most $\tilde{O}(m \log{\delta})$ to evaluate.
These steps combine for a total of
$\tilde{O}(|\Vboundary|m \log(\kappa \epsilon^{-1}) )$.
Also, $\laplacianmiddle^T$ is a sparse matrix with $O(m)$
non-zero entries corresponding to the edges.
So multiplying it against $\matb \laplacianmiddle$, which is
has $|\Vboundary|$ columns can be done
in $O(|\Vboundary|m)$ time.
Combining these two steps gives the overall bound.

\QED

The requirements for \textsc{ApproxSchur} are stated in terms
of spectrum to ease the use of it within recursive routines later
in this section.
We make use of the following lemmas to convert between rough
bounds on edge weights and spectrum.

\begin{lemma}
\label{lem:edgetospectral}
Let $G = (V, E, \weightv)$ be a connected graph with corresponding graph
Laplacian $\laplacian$.
We can find in $O(m)$ time eigenvalue bounds $\lambda_{\min}$
and $\lambda_{\max}$
such that $\lambda_{\min} \leq \lambda_{2}(\laplacian) \leq \lambda_{n}(\laplacian) \lambda_{\max} \leq n^3 \maxratio(\weightv) \lambda_{\min}$. 
\end{lemma}

\Proof
Let $\weight_{\min}$ and $\weight_{\max}$ denote the minimum and maximum
weight in $G$.
Since $G$ is connected, it has a spanning tree consisting of edges of
weight $\weight_{\min}$, giving $\lambda_{\min} \geq \frac{1}{n^2} \weight_{\min}$.

Also, since all edge weights of $G$ are at most $\weight_{\max}$, we have
$\laplacian \preceq \weight_{\max} \laplacian(K_n)$ where $K_n$ is the
complete graph on $n$ vertices.
Therefore $\lambda_{\max} = n \weight_{\max}$ suffices and its relation with
$\lambda_{\min}$ follows from $\weight_{\max} \leq \maxratio \weight_{\min}$.
\QED

\begin{lemma}
\label{lem:spectrallower}
Given a graph $G = (V, E, \weightv)$ and bound
$\lambda_{\min} \leq \lambda_{2}(\laplacian)$.
We can compute in $O(m)$ time another graph $G' = (V, E, \weightv')$
on the same edge set
such that the minimum edge weight in $G'$ is
$\frac{1}{n^2} \lambda_{\min}$ and:
\begin{align*}
\laplacian \preceq \laplacian' \preceq (1 + \frac{1}{n}) \laplacian
\end{align*}

\end{lemma}

\Proof
We have $\frac{\lambda_{min}}{n^2} \laplacian(K_n) \preceq \frac{1}{n} \laplacian$
where $K_n$ is the complete graph on $n$ vertices.
Therefore adding $\frac{\lambda_{\min}}{n^2}$ to each edge of $G$ gives $G'$.
\QED

Once we obtain the approximate inverse, its edge count can be
reduced using spectral sparsifiers.
The guarantees of the sparsification algorithm, first introduced
by Spielman and Srivastava \cite{SpielmanS08} is:
\begin{lemma}
\cite{SpielmanS08, KL11}
\label{lem:sparsify}
There is a routine $\textsc{Sparsify}$ such that given a graph Laplacian
$\laplacian_G$ on $n$ vertices with $m$ edges and a parameter $\epsilon$.
$\textsc{Sparsify}(G, \epsilon)$ outputs in $\tilde{O}(m\log(\epsilon^{-1}))$ time,
a graph Laplacian $\laplacian_H$ with $\tilde{O}(n\epsilon^{-2})$ edges such that
$
(1 - \epsilon) \laplacian_G
\preceq \laplacian_H
\preceq (1+\epsilon) \laplacian_G
$.
\end{lemma}

Combining spectral sparsification with $\textsc{ApproxSchur}$
completes the proof of lemma \ref{lem:onestep}.
The pseudocode of \textsc{OneStepVertexSparsify} is given in Algorithm
\ref{alg:onestepspectralvertexsparsify}

\begin{algorithm}[ht]
\qquad

\textsc{OneStepVertexSparsify}
\vspace{0.05cm}

\underline{Input:}
Graph Laplacian $\laplacian$ corresponding to
$G = (V, E, \weightv) $ with vertices $V$ and boundary vertices $\Vboundary$.
allowable error per step $\epsilon$.

\underline{Output:}
Approximate Schur complement $\laplacianpivotedapprox$.

\begin{algorithmic}[1]
\STATE{Let $\lambda_{\min} = \frac{1}{n^2} \min_{e} \weightv(e)$}
\STATE{Let $\lambda_{\max} = n \max_{e} \weightv(e)$}
\STATE{Let $\maxcond = \lambda_{\max} / \lambda_{\min}$}
\STATE{$\laplacianpivotedapprox \leftarrow
	\textsc{ApproxSchur}(G, \Vboundary, \maxcond, \epsilon/3)$}
\STATE{$\laplacianpivotedapprox' \leftarrow
	\textsc{Sparisfy}(\laplacianpivotedapprox, \epsilon/3)$}
\STATE{Add $\frac{1}{n^2} \lambda_{\min}$ to the weight
of all edges present in $\laplacianpivotedapprox'$ to obtain
$\laplacianpivotedapprox''$}
\RETURN{$\laplacianpivotedapprox''$}
\end{algorithmic}

\caption{One Step Spectral Vertex Sparsification}

\label{alg:onestepspectralvertexsparsify}
\end{algorithm}

\Proofof{Lemma \ref{lem:onestep}}
By Lemma \ref{lem:edgetospectral} we have the
that $\lambda_{\min}$, $\lambda_{\max}$ and $\maxcond$ are
bounds for eigenvalues.
 Lemma \ref{lem:sparsify} gives that
$\laplacianpivotedapprox''$ has $\tilde{O}(n\epsilon^{-2})$ edges and:
\begin{align*}
(1 - \epsilon/3) \laplacianpivotedapprox
\preceq& \laplacianpivotedapprox' \preceq (1 + \epsilon/3)\laplacianpivoted
\end{align*}
Combining with Lemma \ref{lem:approxschurcomplement}
gives that $\laplacianpivotedapprox'$ meets the requirements
of an $\epsilon$-spectral vertex sparsifier.
By Lemmas \ref{lem:approxschur} and \ref{lem:schurextended}, $2\maxcond$ is
sufficient as spectrum bound for $\laplacianpivotedapprox''$.
Therefore using Lemma \ref{lem:spectrallower} we have
$\laplacianapprox'' \preceq (1 + \frac{1}{n}) \laplacianpivotedapprox' 
\preceq  (1 + \epsilon)\laplacianpivoted$.
Furthermore, the edge weights in $\laplacianpivoted''$ are
at least $\frac{1}{n^2} \lambda_{\min}$ and at most
$O(n \lambda_{\max})$, giving a bound of $O(n^5 \maxratio(\weightv))$.
As $\textsc{Sparsify}$ runs in nearly-linear time and $\laplacianpivotedapprox$
has at most $|\Vboundary|^2 \leq |\Vboundary|m$ edges,
$\textsc{ApproxSchur}$ is the bottleneck in the total run time.
\QED

We now move on to our multilevel recursive sparsification algorithm for graphs
where we have access to a full separator tree.
Pseudocode of our algorithm is shown in Algorithm \ref{alg:recursivespectralvertexsparsify}

\begin{algorithm}[ht]
\qquad

\textsc{VertexSparsify}
\vspace{0.05cm}

\underline{Input:}
Node in separator tree $\mathcal{S}$, graph $G$, with separator $S$ and
children  $\component{1} = (V^{(1)}, E^{(1)})$ and
$\component{2} = (V^{(2)}, E^{(2)})$.
Graph Laplacian $\laplacian$ corresponding to
$G$ with vertices $V$ and boundary vertices $\Vboundary$.
Spectrum bound $\maxcond$ such that $\lambda_{n}(\laplacian)\
\leq \maxcond \lambda_{2}(\laplacian)$,
allowable error per step $\epsilon$.

\underline{Output:}
Approximate Schur complement $\laplacianpivotedapprox$.

\begin{algorithmic}[1]
\FOR{$i = \{1, 2\}$}
	\STATE{$\VboundaryG{G^{(i)}} \leftarrow (V^{(i)} \cap \Vboundary) \cup S$}
	\STATE{$\laplacianpivotedapprox^{(i)} \leftarrow \textsc{VertexSparsify}
		(\laplacian^{(i)}, \VboundaryG{G^{(i)}}, \maxcond, \epsilon)$}
\ENDFOR
\STATE{$\laplacianpivotedapprox' \leftarrow \laplacianpivotedapprox^{(1)}
	+ \laplacianpivotedapprox^{(2)}$} \label{ln:laplacianpivotedapprox1}
\STATE{$\laplacianpivotedapprox'' \leftarrow \textsc{ApproxSchur}
	(\laplacianpivotedapprox', \Vboundary, 	
			2^{\log_{20/19}{n}} \maxcond, \epsilon / 3)$}
\STATE{$\laplacianpivotedapprox \leftarrow
	\textsc{Sparsify}(\laplacianpivotedapprox'', \epsilon / 3)$}
\RETURN{$\laplacianpivotedapprox$}
\end{algorithmic}

\caption{Recursive Spectral Vertex Sparsification for Well-Separated Graphs}

\label{alg:recursivespectralvertexsparsify}
\end{algorithm}

In order to provide error guarantees for \textsc{VertexSparsify},
we relate rough spectral bounds between.
This in turn allows applications of Lemma
\ref{lem:approxschurcomplement} in an inductive fashion.
These guarantees can be summarized as follows:

\begin{lemma}
\label{lem:recursivespectralvertexsparsify}
Let $G$ be a graph with $n$ vertices and $m$ edges such that
$\lambda_{n}(G) \leq \kappa \lambda_{2}(G)$,
Given a partition of the vertices $V$ into boundary and interior
vertices such that $|\Vboundary| \leq \sqrt{n}$,
a $9/10$ separator tree $\mathcal{S}$,
and an error parameter $\epsilon$.

Then $\laplacianpivotedapprox'$ as defined on Line
\ref{ln:laplacianpivotedapprox1} of \textsc{VertexSparsify} satisfies:
\begin{align*}
\lambda_n(\laplacianpivotedapprox')
\leq & 2^{\log_{20/19}{n}} \maxcond \lambda_2(\laplacianpivotedapprox')
\end{align*}
And $\textsc{VertexSparsify}(\laplacian, \Vboundary, \maxcond, \epsilon)$
returns $\laplacianpivotedapprox$ with $\tilde{O}(n\epsilon^{-2})$ edges
such that:
\begin{align*}
(1 - \epsilon)^{\log_{20/19}{n}} \laplacianpivoted \preceq
\laplacianpivotedapprox
\preceq (1 + \epsilon)^{\log_{20/19}{n}} \laplacianpivoted
\end{align*}
\end{lemma}

\Proof
The proof is by induction on $n$.
Since $|S|$ has size at most $O(\sqrt{n})$ we have that
$|\VboundaryG{\component^{1}}|, |\VboundaryG{\component{2}}| \leq O(\sqrt{n})$.
Then since the two graphs that we recurse on have size at most
$9/10 n + O(\sqrt{n}) \leq 19/20 n$, by the induction hypothesis we have:
\begin{align*}
(1 - \epsilon)^{\log_{20/19}{n} - 1} \laplacianpivoted^{(i)} \preceq
\laplacianpivotedapprox^{(i)}
\preceq (1 + \epsilon)^{\log_{20/19}{n} - 1} \laplacianpivoted^{(i)}
\end{align*}

Furthermore, we can compute an approximate Schur complement
of this sum,  only introducing an extra
error of $(i - 1) / 2\log{n} \epsilon $
Let $\laplacianpivoted'$ be the (exact) Schur complement
of $(G, \Vinterior \setminus S)$.
Then invoking the induction hypothesis and summing over the two
pieces gives:
\begin{align*}
(1 - \epsilon)^{\log_{20/19}{n} - 1} \laplacianpivoted' \preceq
\laplacianpivotedapprox'
\preceq (1 + \epsilon)^{\log_{20/19}{n} - 1} \laplacianpivoted'
\end{align*}

Combining this with the relation of the spectrum of $\laplacianpivoted$
and $\laplacian$ gives:
\begin{align*}
& \lambda_n(\laplacianpivotedapprox') \nonumber \\
\leq & (1 + \epsilon)^{\log_{20/19}{n}} \lambda_n(\laplacianpivoted) \nonumber \\
\leq & (1 + \epsilon)^{\log_{20/19}{n}} n \lambda_n(\laplacian) \nonumber \\
\leq & (1 + \epsilon)^{\log_{20/19}{n}} n
      \maxcond \lambda_2(\laplacian) \nonumber \\
\leq & (1 + \epsilon)^{\log_{20/19}{n}} (1 - \epsilon)^{-\log_{20/19}{n}}
 	n \maxcond \lambda_2(\laplacianpivotedapprox') \nonumber \\
\leq & 2^{\log_{20/19}{n}} n \maxcond \lambda_2(\laplacianpivotedapprox')
\nonumber \\ & 
\qquad \text{By assumption that $\epsilon \leq 1/10$, }
\end{align*}

Therefore running $\textsc{ApproxSchur}$ with
spectrum bound of $2^{\log_{20/19}{n}} n \maxcond$
satisfies the input requirements for in Lemma \ref{lem:approxschur}.
Let $\laplacianpivoted''$ in turn be the exact Schur complement of
removing $S$ from $\laplacianpivotedapprox'$,
Lemma \ref{lem:approxschur} then gives that the approximation
guarantee still holds.
\begin{align*}
(1 - \epsilon)^{(\log_{20/19}{n}) - 1} \laplacianpivoted \preceq
\laplacianpivoted''
\preceq (1 + \epsilon)^{(\log_{20/19}{n}) - 1} \laplacianpivoted
\end{align*}

By Lemma \ref{lem:approxschurcomplement},
$\laplacianpivotedapprox''$ satisfies:
\begin{align*}
(1 - \epsilon/3) \laplacianpivoted'' \preceq
\laplacianpivotedapprox''
\preceq (1 + \epsilon/3) \laplacianpivoted''
\end{align*}

Lemma \ref{lem:sparsify} gives an additional error
of $1 \pm \epsilon / 3$ from the sparsification, as well
as the bound on edge count.
Including this error gives:
\begin{align*}
(1 - \epsilon) \laplacianpivoted'' \preceq
\laplacianpivotedapprox
\preceq (1 + \epsilon)\laplacianpivoted''
\end{align*}
Combining this with the bound between $\laplacianpivoted$
and $\laplacianpivoted''$ shows the inductive
hypothesis for $n$ as well.
\QED

We now turn our attention to the running time, which is dominated
by the calls to $\textsc{ApproxSchur}$.

\begin{lemma}
\label{lem:recursivespectralvertexsparsifytime}
Let $G$ be a graph with $n$ vertices and $m$ edges such that
$\lambda_{n}(G) \leq \kappa \lambda_{2}(G)$,
Given a partition of the vertices $V$ into boundary and interior
vertices such that $|\Vboundary| \leq \sqrt{n}$,
a $9/10$ separator tree $\mathcal{S}$,
and an error parameter $\epsilon$,
$\textsc{VertexSparsify}(\laplacian, \Vboundary, \maxcond, \epsilon)$
terminates in $\tilde{O}((m + |\Vboundary|^2)\epsilon^{-2}\log(\maxcond))$ time
\end{lemma}

\Proof

Since $\laplacianpivotedapprox^{(1)}$ and $\laplacianpivotedapprox^{(1)}$
combine for $|\VboundaryG{\component{(1)}}| +  |\VboundaryG{\component{2}}|
\leq 2(|\Vboundary| + |S|)$ vertices, $\laplacianpivotedapprox'$
has at most $\tilde{O}((|\Vboundary| + |S|) \epsilon^{-2})$ edges by
the induction hypothesis.
So by Lemma \ref{lem:approxschurcomplement}, the running time of this
call is bounded by $\tilde{O}((|\Vboundary| + |S|)|\Vboundary| \epsilon^{-2}
\log(\maxcond))$.

Therefore the total running time can be bounded by
$\tilde{O}(\epsilon^{-2} \log(\maxcond))$
times the total value of $(|\Vboundary| + |S|)|\Vboundary|$
summed over all the recursive calls.
We follow the proof of Theorem 2 of \cite{LiptonRT79},
which bounds the fill of generalized nested dissection.
We also and use $f(l, n)$ to denote the maximum
sum of $(|\Vboundary| + |S|)|\Vboundary|$ over the recursive
calls starting from a graph with
$n$ vertices, $l$ of which are on the boundary.

Let $c_0$ be the constant in the $9/10$ separator tree
where the separator size is at most $c_0 \sqrt{n}$.
Then the recursive calls guarantee that:

\begin{itemize}

\item If $n \leq n_0$ for some absolute constants $C$ and $n_0$, then $f(l, n) \leq C$.

\item Otherwise, $f(l, n) \leq l (l + c_0 n) + f(l_1, n_1) + f(l_2, n_2)$ 
for  $l_1, l_2, n_1, n_2$ that satisfy:
\begin{align*}
l_1 + l_2 \leq & l + 2 c_0 \sqrt{n}\\
n_1 + n_2 \leq & n + c_0 \sqrt{n}\\
n_1, n_2 \leq & 9/10 n + c_0 \sqrt{n}
\end{align*}

\end{itemize}

Then we show by induction on $n$ that there exists an absolute
constants $c_1$ and $c_2$ such that:
\begin{align*}
f(l, n) \leq c_1 l^2 \log{n} + c_2 n \log^2 {n}
\end{align*}

The base case of $n \leq n_0$ follows from $C$ being a constant.
For the inductive case the RHS is bounded by:
\begin{align*}
& c_1 l_1^2 \log{n_1} + c_2 n_1 \log^2{n_1}
\nonumber \\ & ~~
+ c_1 l_2^2 \log_{n2}+ c_2 n_2 \log^2 {n_2} +  l(l + c_0 \sqrt{n})
\nonumber \\
\leq & c_1 \log(\max(n_1, n_2)) (l_1^2 + l_2^2)
\nonumber \\ & ~~
+ c_2 \log(\max(n_1, n_2))  (n_1 + n_2) + l(l + c_0 \sqrt{n}) \nonumber \\
\leq & c_1 (\log(n) - \log(10/9) )(l_1 + l_2)^2
\nonumber \\ & ~~
+ c_2 (\log(n) - \log(10/9))^2  (n_1 + n_2) + l(l + c_0 \sqrt{n}) \nonumber \\
\leq & c_1 (\log(n) - \log(10/9) )(l + 2 c_0 \sqrt{n})^2
\nonumber \\ & ~~
+ c_2 (\log(n) - \log(10/9))^2  (n + c_0 \sqrt{n})
+ l(l + c_0 \sqrt{n}) \nonumber \\
\end{align*}

By the arithmetic-geometric mean inequality we have $l \cdot c_0 \sqrt{n}
\leq (l^2 + c_0^2 n)/2$.
Making this substitution gives:
\begin{align*}
\leq & c_1 (\log(n) - \log(10/9) )(l^2 + 8 c_0^2 n)
\nonumber \\ & ~~
+ c_2 (\log(n) - \log(10/9))^2  (n + c_0 \sqrt{n})
+ 2l^2 + c_0^2 n \nonumber \\
\leq & (c_1 \log(n) - c_1 \log(10/9) + 2) l^2
\nonumber \\ & ~~
+ \left( c_2 (\log(n) - \log(10/9))^2 + 8 c_1 \log(n) c_0^2 + c_0^2 \right) n
\nonumber \\ & ~~
+ c_2 \log^2(n) c_0 \sqrt{n} \nonumber
\end{align*}

By an appropriate choice of $n_0$ we can ensure that
$\log^2(n) \sqrt{n} \leq n$, giving a simpler upper bound of:
\begin{align*}
\leq & (c_1 \log(n) - c_1 \log(10/9) + 2) l^2
\nonumber \\ & ~~
+ \left( c_2 (\log(n) - \log(10/9))^2 + 8 c_1 \log(n) c_0^2 + 2 c_0^2 \right) n
\end{align*}

By choosing $c_1 = 2 / \log(10/9)$, we get the coefficient on $l^2$
to be at most $c_1 \log{n}$.
Then since $c_2 \log(n)^2 -  c_2 (\log(n) - \log(10/9))^2
= 2 c_2 \log(10/9) - \log(10/9)^2$, and $c_1$, $c_0$ are fixed
constants, a sufficiently large constant for $c_2$ ensures
the coefficient in front of $n$ is also bounded by $c_2 \log{n}$.
Combining with the extra term of $\tilde{O}(\epsilon^{-2} \log(\maxcond))$
gives the first term in the running time bound.
An additional term of $m$ is needed to account for the cost of the initial
sparsification.

\QED

\Proofof{Lemma \ref{lem:recursive}}
The conversion between guarantees on edge weight ratio
$\maxratio(\weightv)$ and spectrum can be handled
in the same way as in Lemma \ref{lem:onestep}.
Combining Lemmas \ref{lem:recursivespectralvertexsparsify} and
\ref{lem:recursivespectralvertexsparsifytime},
and adjusting parameters by replacing $\epsilon$ with
$\epsilon / 2\log_{20/19}{n}$
gives the required bounds.
\QED

\section{Proofs of Properties of Schur Complement}
\label{sec:schurproofs}

\Proofof{Lemma \ref{lem:pivotok}}

Let the vectors $\vecx$ be the solution to the solution to
$\laplacian \vecx = \demandv$.
Partitioning $\vecx$ into $\vecxboundary$ and $\vecxinterior$ gives:
\begin{align*}
\left [
\begin{array}{cc}
\laplacianinner & \laplacianmiddle \\
\laplacianmiddle^T & \laplacianboundary\\
\end{array}
\right]
\left[
\begin{array}{c}
\vecxinterior\\
\vecxboundary\\
\end{array}
\right]
& =
\left[
\begin{array}{c}
\zerosv\\
\demandboundaryv\\
\end{array}
\right]
\end{align*}

The equality given by the first set of blocks can be rewritten as:
\begin{align*}
\laplacianinner \vecxinterior + \laplacianmiddle \vecxboundary =& \zerosv \nonumber \\
\vecxinterior = & - \laplacianinner^{-1} \laplacianmiddle \vecxboundary
\end{align*}

Substituting this into the second set of blocks gives:
\begin{align*}
-\laplacianmiddle^T \laplacianinner^{-1} \laplacianmiddle \vecxboundary
+ \laplacianboundary \vecxboundary
= & \demandboundaryv \nonumber \\
(\laplacianboundary - \laplacianmiddle^T \laplacianinner^{-1} \laplacianmiddle)
\vecxboundary = & \demandboundaryv \nonumber \\
\vecxboundary = & \laplacianpivoted \demandboundaryv
\end{align*}

Which if we substitute into $\demand^T \laplacian^+ \demandv$
and combining with the fact that $\demandv$ is only non-zero on the boundary
vertices gives:
\begin{align*}
\demandv^T \laplacian^+ \demandv
= & \demandboundaryv^T \vecxboundary
= \demandboundaryv^T \laplacianpivoted^{+} \demandboundaryv(i)
\end{align*}
\QED{Lemma \ref{lem:pivotok}}

\Proofof{Lemma \ref{lem:schurextended}}

If we treat $\vecx$ as a constant, then the expression to be minimized
can be written as:
\begin{align*}
& \vecyx^T
\laplacian(G)
\vecyx \nonumber \\
= &
\vecyx^T
\left [
\begin{array}{cc}
\laplacianinner & \laplacianmiddle \\
\laplacianmiddle^T & \laplacianboundary\\
\end{array}
\right]
\vecyx \nonumber \\
= & \vecy^T \laplacianinner \vecy + 2 \vecy^T \laplacianmiddle \vecx
+ \vecx^T \laplacianboundary \vecx
\end{align*}

The first two terms can be viewed as a quadratic in $\vecy$,
and is therefore minimized at $\bar{y} = -\laplacianinner^{-1} \laplacianmiddle \vecx$.
Substituting this into the expression above gives:
\begin{align*}
& \min_{y} 
\vecyx^T
\laplacian(G)
\vecyx \nonumber \\
= & \bar{\vecy}^T \laplacianinner \bar{\vecy} + 2 \bar{\vecy}^T \laplacianmiddle \vecx
+ \vecx^T \laplacianboundary \vecx \nonumber \\
= & \vecx^T \laplacianboundary \vecx - \vecx^T \laplacianmiddle^T \laplacianinner^{-1} \laplacianmiddle \vecx \nonumber \nonumber \\
= & \vecx^T (\laplacianboundary -  \laplacianmiddle^T \laplacianinner^{-1} \laplacianmiddle) \vecx \nonumber \\
= & \vecx^T \laplacianpivoted \vecx
\end{align*}
\QED{Lemma \ref{lem:schurextended}}

\Proofof{Lemma \ref{lem:schurspectrum}}

Let $\vecx$ be any vector on the support of $\laplacianpivoted$.
Then by Lemma \ref{lem:schurextended} there exist a vector $\vecy$
such that:
\begin{align}
\vecx^T \laplacianpivoted(G) \vecx
=&
\vecyx^T
\laplacian(G)
\vecyx
\end{align}

Furthermore, all entries of $\vecy$ are between the minimum and
maximum entries of $\vecx$, giving:
\begin{align}
\vecx^T \vecx
\leq & \vecx^T \vecx + \vecy^T\vecy
\leq n \vecx^T \vecx
\end{align}

The bounds then follow by setting $\vecx$ to eigenvectors of $\laplacianpivoted$
and applying the Courant-Fisher theorem on $\laplacian$.
\QED

\section{Returning a Small Cut}
\label{sec:cut}

We now show that if \textsc{GroupedFlow} returns \fail, then a small cut can
also be found.
We begin by stating one omitted aspect of the electrical flow algorithm as
given in Theorem 2.3 of \cite{ChristianoKMST10}.

\begin{lemma}
\label{lem:voltages}
\textsc{ElectricalFlow} can be modified to return a set of vertex potentials
$\approxpotentialvertv$ such that the energy given by the potentials
$\epotential_{\approxpotentialvertv}(\resistancev)$ is at most $1$,
and $\demandv^T \approxpotentialvertv$ is at least $(1 - \delta) \eflow(\optflow)$.
\end{lemma}

These vertex potentials are approximate versions of the voltages corresponding
to the electrical flow.
Since the vertex sparsified graph is approximately the same as the graph given
as input to \textsc{GroupedFlow}, we show that this can be extended to
a full set of potentials on vertices of the original graph, which in turn leads
to a small cut.

\begin{theorem}
If $\textsc{GroupedFlow}$ returns \fail, we can find in $\tilde{O}(m)$ time a
set of vertex potentials $\potentialvertv$ such that:

\begin{enumerate}
\item $\sum_{e = uv \in E} \capacity(e) |\potentialvert_u - \potentialvert_v| \leq 1$
\item $\demandv^T \potentialvertv \geq 1 - 10\epsilon$
\end{enumerate}

\end{theorem}

\Proof

By Lemma \ref{lem:voltages}, in this situation
a set of vertex potentials $\approxpotentialschurv$ on $\Gschurapprox$ such that:
$\approxpotentialschurv^T \laplacianpivotedapprox \approxpotentialschurv
\leq 1$
and $\demandboundaryv^T \approxpotentialschurv \geq \mu$
where $\mu = \sum_{1 \leq i \leq k} \weightgroup(i)$.
Let the weights among edges given \textsc{Oracle} be $\weightv$
and the weights set in \textsc{GroupedFlow} be $\weightgroupv$.
Furthermore, for $e \in E(\component{i})$ we use $\weightcombined(e)$ to denote
combined weight on $e$ that does not account for its capacity:

\begin{align*}
\weightcombined(e)
= & \capacity(e)^2 \resistance(e)  \nonumber \\
= & (\weightgroup(i) + \frac{\epsilon}{k} \mu) \cdot \left(  \frac{\weight(e)}{\weight(\component{i})} + \frac{\epsilon}{|E(\component{i})|} \right)
\end{align*}

Consider each component $i$, let $\approxpotentialschurv(i)$ denote the restriction
of $\approxpotentialschurv$ to the vertices on $\VboundaryG{\component{i}}$.
Since $\laplacianpivotedapprox(\component{i}) \preceq (1 + \epsilon) \laplacianpivoted(\component{i}) $, we have:

\begin{align*}
\approxpotentialschurv(i)^T \laplacianpivoted(\component{i}) \approxpotentialschurv(i)
\leq (1 + \epsilon) \approxpotentialschurv(i)^T
	\laplacianpivotedapprox(\component{i}) \approxpotentialschurv(i)
\end{align*}

Therefore by Lemma \ref{lem:schurextended},
it's possible to extend $\approxpotentialschurv(i)$ onto the interior vertices
of $\component{i}$ to obtain $\approxpotentialvertv(i)$ such that:

\begin{align*}
\approxpotentialvertv(i)^T \laplacian(\component{i}) \approxpotentialvertv(i)
\leq (1 + \epsilon)  \approxpotentialschurv(i)^T \laplacianpivotedapprox(\component{i}) \approxpotentialschurv(i)
\end{align*}

Summing over all clusters gives a vector $\approxpotentialvertv$ such that:

\begin{align}
\approxpotentialvertv^T \laplacian \approxpotentialvertv
= & \sum_{1 \leq i \leq k} \approxpotentialvertv(i)^T \laplacian(\component{i}) \approxpotentialvertv(i) \nonumber \\
\leq & \sum_{1 \leq i \leq k} \approxpotentialschurv(i)^T \laplacianpivotedapprox(\component{i}) \approxpotentialschurv(i) \nonumber \\
= &(1 + \epsilon) \approxpotentialschurv^T \laplacianpivotedapprox(\component{i}) \approxpotentialschurv \nonumber \\
\leq & 1 + \epsilon
\label{eq:totalpotentialpivoted}
\end{align}

Furthermore since $\demandv$ is $0$ in all interior vertices, we have:

\begin{align*}
\demandv^T \approxpotentialvertv
= & \demandboundaryv^T \approxpotentialschurv \geq \mu
\end{align*}

Then consider $\potentialvertv = \frac{1}{(1+10\epsilon) \mu}$.
We have $\demandv^T \potentialvertv \geq 1 - 10\epsilon$,
and from Equation \ref{eq:totalpotentialpivoted} we can obtain:

\begin{align}
\potentialvertv^T \laplacian \potentialvertv
\leq & \frac{1 + \epsilon}{(1+10\epsilon)\mu} \nonumber \\
(1+8\epsilon) \mu
(\potentialvertv^T \laplacian \potentialvertv)
\leq & 1
\label{eq:totalpotential}
\end{align}

Substituting in the fact that the resistive weight of edge
$e \in E(\component{i})$ in $\laplacian$
is $\resistance(e) = \frac{\weightcombined(e)}{\capacity(e)^2}$ gives:

\begin{align*}
\potentialvertv^T \laplacian \potentialvertv
= & \sum_{1 \leq i \leq k} \sum_{e=uv \in E(\component{i})}
\frac{(\potentialvertv(u) - \potentialvertv(v))^2}{\resistance(e)} \nonumber \\
= & \sum_{e=uv \in E(\component{i})}
\frac{\capacity(e)^2 (\potentialvertv(u) - \potentialvertv(v))^2}
{ \weightcombined(e)}
\end{align*}

Also, we have that the total sum of resistive values among all the
clusters is:

\begin{align*}
& \sum_{e} \weightcombined(e) \nonumber \\
= & \sum_{1 \leq i \leq k} \sum_{e \in E(\component{i})}
(\weightgroup(i)  + \frac{\epsilon}{k} \mu) \cdot \left(  \frac{\weight(e)}{\weight(\component{i})} + \frac{\epsilon}{|E(\component{i})|} \right)
 \nonumber \\
= & \sum_{1 \leq i \leq k} (\weightgroup(i)  + \frac{\epsilon}{k} \mu) \left(
	\sum_{e \in E(\component{i})}  \frac{\weight(e)}{\weight(\component{i})} + \frac{\epsilon}{|E(\component{i})|} \right) \nonumber\\
= & (1  + \epsilon) \sum_{1 \leq i \leq k}
	(\weightgroup(i)  + \frac{\epsilon}{k} \mu) \nonumber \\
\leq & (1 + 3\epsilon) \mu
\end{align*}

Substituting into Equation \ref{eq:totalpotential} gives:

\begin{align*}
1
\geq &
(1+8\epsilon) \mu
(\potentialvertv^T \laplacian \potentialvertv) \nonumber \\
\geq & \left( \sum_{e} \weightcombined(e) \right)
\left( \sum_{e} \frac{\capacity(e)^2 (\potentialvertv(u) - \potentialvertv(v))^2}
{\weightcombined(e)} \right) \nonumber \\
\geq & \left( \sum_{e} \capacity(e) |\potentialvertv(u) - \potentialvertv(v)| \right)^2
\nonumber \\ & \qquad
\qquad \text{By the Cauchy-Schwarz inequality}
\end{align*}

Taking square root of both sides completes the proof.

\QED

\end{appendix}

\end{document}